# A new view on superfluidity

V.A. Golovko

Moscow State Evening Metallurgical Institute, Lefortovsky Val 26, Moscow 111250, Russia

E-mail: mgvmi-mail@mtu-net.ru

## Abstract

This paper represents the full version of a paper published earlier in Physica A [**246** (1997), 275]. The present paper includes argumentation, proofs and details omitted in the shortened version. The papers are a further development of the approach in quantum statistical mechanics proposed by the author. The hierarchy of equations for reduced density matrices obtained previously is extended to the case corresponding to the Bose-Einstein condensation. The relevant state of the system with a condensate can be superfluid as well as nonsuperfluid. Special attention is given to the thermodynamics of superfluid systems. According to the results of the papers superfluidity is the state of a fluid whose symmetry is spontaneously broken because of a stationary flow. The state corresponds to thermodynamic equilibrium while the magnitude of the flow depends upon the temperature and is determined by thermodynamic considerations. The equations obtained are solved in two simple cases. The physical origin of superfluidity, peculiarities of the phenomenon in closed volumes and the critical velocity are discussed as well.

# 1. Introduction

In a previous paper [1] (hereafter referred to as I) it was shown that thermodynamic and structural properties of an equilibrium quantum system can be treated on the basis of a hierarchy of equations for *s*-particle reduced density matrices, the hierarchy being derived in the same paper. In the event of a Bose system, however, the examples considered in I show that the equations obtained have no solution at low temperatures, which points out that special investigation is needed in order to extend the approach of I to that temperature region while bearing in mind that any correct physical theory of many-body Bose systems must necessarily lead to such a phenomenon as superfluidity. The present paper, which is a continuation of the work begun in I, deals with the low-temperature behaviour of a Bose system and especially with superfluidity.

It should be emphasized that as yet there is no physically convincing explanation for the phenomenon of superfluidity. Superfluidity is customarily explained on a base of Landau's argument [2,3] which, being purely classical in essence (in the argument figure neither the Planck constant nor the type of statistics), is applied to a highly quantum phenomenon. Moreover the Landau criterion yields critical velocities that are substantially higher than those really observed. It should be added that in no way does Landau's explanation relate superfluidity with the Bose-Einstein condensation whereas such a relation was supposed soon after the discovery of the phenomenon [4] and is ascertained experimentally [5].

Feynman [6] tried to explain the discrepancy between the Landau criterion and experiment by formation of vortices. However Feynman's reasoning, refuting in fact Landau's one, does not explain the phenomenon of superfluidity (from Feynman's reasoning by itself it does not follow the existence of a nonzero critical velocity), nor does it relate the phenomenon with the Bose-Einstein condensation either. Besides, from Feynman's argument as well as from Landau's one it is not at all clear why the critical velocities tend to zero as the $\lambda$ point is approached. Other theories of superfluidity such as Bogolyubov's one [7] come down in effect to a proof that the spectrum of elementary excitations corresponds with Landau's argument while with this argument corresponds any acoustic-type spectrum.

The absence of a consistent theory of superfluidity as an equilibrium phenomenon led some authors to the conclusion that the superfluid state is not equilibrium but metastable [8-10]. Lee and Yang [11] (see Ref. [12] as well) also considered superfluidity to be a nonequilibrium phenomenon and argued that the viscosity should vanish only at zero temperature. Sometimes even the term "superfluidity" is not used in the initial sense, but is used for indicating off-diagonal long range order (ODLRO) [13]. We shall understand superfluidity in the initial sense of the word, as an equilibrium state of the fluid in which there is a persistent flow without any dissipation of energy.

In the present paper we shall lean on the phenomena that occur in the condensed phase of an ideal Bose gas, and we shall see that interaction between particles of the system can give rise to superfluidity. It should be stressed that this study is based entirely upon the ideas developed in I, and there is no need to resort to any additional hypotheses or to make use of an argument similar to the ones mentioned above. In this paper, first of all we shall find out the point in the argumentation of I which should be modified in order to include superfluidity into the theory. Next, following the scheme of I we shall obtain a modified hierarchy of equations, consider thermodynamics and provide examples of solution of the equations. In the concluding section we shall discuss physical causes explaining superfluidity and broach some other aspects of the problem as well.

In the present paper we consider the same systems of particles as in I (mainly Bose systems, however) and use the same notation. For the sake of convenience, when referring to an equation of I, we shall place I in front; so, for example, Eq. (I.2.7) will stand for Eq. (2.7) of I.

## 2. Basic equations

The starting point of I is the hypothesis that at thermodynamic equilibrium the reduced density matrices (we imply the term in Eq. (I.2.7) which does not depend upon the time) are of the form (see Eq. (I.2.11))

$$R_s(\mathbf{x}_s, \mathbf{x}'_s) = \sum_\nu n_s\left(\varepsilon_\nu^{(s)}\right) \psi_\nu(\mathbf{x}_s) \psi_\nu^*(\mathbf{x}'_s), \qquad (2.1)$$

where $\nu$ represents a set of numbers that determine the eigenfunctions $\psi_\nu(\mathbf{x}_s)$ of Eq. (I.2.8). In I it was assumed that the functions $n_s(z)$ (here $z = \varepsilon_\nu^{(s)}$) were smooth and even analytic.

However, the smoothness of $n_s\left(\varepsilon_\nu^{(s)}\right)$ is not obligatory in the hypothesis expressed by (2.1). Let us suppose now that $n_s\left(\varepsilon_\nu^{(s)}\right)$ is mainly a smooth function of $\varepsilon_\nu^{(s)}$ except for a set $\nu = \nu_0$ (depending on $s$) at which its magnitude is larger than that required by the smoothness. Then Eq. (2.1) can be rewritten as

$$R_s(\mathbf{x}_s, \mathbf{x}'_s) = R_s^{(c)}(\mathbf{x}_s, \mathbf{x}'_s) + R_s^{(n)}(\mathbf{x}_s, \mathbf{x}'_s) \qquad (2.2)$$

with

$$R_s^{(c)}(\mathbf{x}_s, \mathbf{x}'_s) = \varphi_s(\mathbf{x}_s) \varphi_s^*(\mathbf{x}'_s), \qquad R_s^{(n)}(\mathbf{x}_s, \mathbf{x}'_s) = \sum_\nu n_s\left(\varepsilon_\nu^{(s)}\right) \psi_\nu(\mathbf{x}_s) \psi_\nu^*(\mathbf{x}'_s), \qquad (2.3)$$

where the superscript (*c*) means the condensate fraction and (*n*) the normal one (we resort to the terminology used in consideration of the Bose condensation of an ideal gas). In $R_s^{(n)}$ of (2.3), we have retained the term with $\nu = \nu_0$ upon smoothing down the function $n_s\left(\varepsilon_\nu^{(s)}\right)$. Because of this the term $R_s^{(c)}$ appears in which the function $\varphi_s(\mathbf{x}_s)$ differs from $\psi_{\nu_0}(\mathbf{x}_s)$ in the normalization only: the function $\psi_{\nu_0}(\mathbf{x}_s)$ is normalized to unity as in I while the normalization of $\varphi_s(\mathbf{x}_s)$ is unknown for the moment since in its normalization is included the difference between the initial and smoothed magnitudes of $n_s\left(\varepsilon_{\nu_0}^{(s)}\right)$. It should be observed that, if the initial magnitude of $n_s\left(\varepsilon_{\nu_0}^{(s)}\right)$ were smaller than that required by the smoothness, the sign of $R_s^{(c)}$ should be reversed; this case, however, will not be considered.

If several states are exceptional, we shall have the following expression, instead of (2.2) and (2.3),

$$R_s(\mathbf{x}_s, \mathbf{x}'_s) = \sum_i \varphi_s^{(i)}(\mathbf{x}_s)\,\varphi_s^{(i)*}(\mathbf{x}'_s) + \sum_\nu n_s\left(\varepsilon_\nu^{(s)}\right)\psi_\nu(\mathbf{x}_s)\psi_\nu^*(\mathbf{x}'_s). \qquad (2.4)$$

As long as the functions $n_s(z)$ in (2.3) and (2.4) are smooth, Eqs. from (I.2.16) to (I.2.24) remain valid for $R_s^{(n)}(\mathbf{x}_s, \mathbf{x}'_s)$. As to the function $\varphi_s(\mathbf{x}_s)$, according to (I.2.8) it obeys the equation

$$\frac{\hbar^2}{2m}\sum_{j=1}^s \nabla_j^2 \varphi_s(\mathbf{x}_s) + [\varepsilon_{(s)} - U_s(\mathbf{x}_s)]\,\varphi_s(\mathbf{x}_s) = 0, \qquad (2.5)$$

in which $\varepsilon_{(s)}$ is written for $\varepsilon_{\nu_0}^{(s)}$. When deducing Eqs. (I.2.14) and (I.3.1) for the effective potentials $U_s(\mathbf{x}_s)$ no condition on $n_s\left(\varepsilon_\nu^{(s)}\right)$ was implied, therefore they can be used as they stand. Analogously with (2.2) we write for the diagonal elements

$$\rho_s(\mathbf{x}_s) = \rho_s^{(c)}(\mathbf{x}_s) + \rho_s^{(n)}(\mathbf{x}_s), \quad \rho_s^{(c)}(\mathbf{x}_s) = |\varphi_s(\mathbf{x}_s)|^2 \qquad (2.6)$$

with Eq. (I.3.3) for $\rho_s^{(n)}(\mathbf{x}_s)$.

Thus, as in I we have again obtained a hierarchy of equations for the diagonal elements of the reduced density matrices, which contains $\rho_s(\mathbf{x}_s)$, $U_s(\mathbf{x}_s)$, $v_s(\mathbf{x}_s, \mathbf{m}_s, z)$, and now $\varphi_s(\mathbf{x}_s)$ in addition. Let us write down the equations of the hierarchy (we retain the upper sign in (I.3.3) referring to bosons):

$$\rho_s(\mathbf{x}_s) = |\varphi_s(\mathbf{x}_s)|^2 + \frac{1}{2\pi i(2\pi\hbar)^{3s}s!}\int d\mathbf{m}_s \int_C dz\, n_s(z)\, v_s(\mathbf{x}_s,\mathbf{m}_s,z)\sum_P^{(s)}\exp\left[\frac{i}{\hbar}\sum_{k=1}^s \mathbf{r}_k(\mathbf{p}_k - \mathsf{P}\mathbf{p}_k)\right], (2.7)$$

$$\frac{\hbar^2}{2m}\sum_{j=1}^{s}\nabla_j^2\, v_s + \frac{i\hbar}{m}\sum_{j=1}^{s}\mathbf{p}_j\nabla_j\, v_s + \left[z - \frac{1}{2m}\sum_{j=1}^{s}\mathbf{p}_j^2 - U_s(\mathbf{x}_s)\right] v_s = 1, \qquad (2.8)$$

$$\rho_s(\mathbf{x}_s)\,\nabla_1 U_s(\mathbf{x}_s) = \rho_s(\mathbf{x}_s)\,\nabla_1\left[\sum_{j=2}^{s} K(|\mathbf{r}_1-\mathbf{r}_j|)+V^{(e)}(\mathbf{r}_1)\right] + \int \rho_{s+1}(\mathbf{x}_{s+1})\nabla_1 K(|\mathbf{r}_1-\mathbf{r}_{s+1}|)d\mathbf{r}_{s+1}, \quad (2.9)$$

the fourth equation being (2.5). The quantities $n_s(z)$, $\varepsilon_{(s)}$ and the normalization of $\varphi_s(\mathbf{x}_s)$ are arbitrary in this hierarchy.

In order to find those quantities we turn to the interrelation of (I.4.1). By virtue of its linearity, we can require it to be satisfied by $R_s^{(c)}$ and $R_s^{(n)}$ separately. Seeing that $R_s^{(n)}$ obeys the same equations as $R_s$ in I, following the procedure of Sec. 4 of I we arrive at Eq. (I.4.11). To obtain an additional condition on $n_s(z)$ we may use the same argument as in I. Let our system be formed by two mutually noninteracting subsystems $A$ and $B$. In this case the density matrices must be of the form

$$R_s(\mathbf{x}_s, \mathbf{x}_s') = C R_{s_a}\!\left(\mathbf{x}_{s_a}^{(a)}, \mathbf{x}_{s_a}^{(a)'}\right) R_{s_b}\!\left(\mathbf{x}_{s_b}^{(b)}, \mathbf{x}_{s_b}^{(b)'}\right), \qquad (2.10)$$

where the constant $C$ depends on the manner in which one normalizes $R_{s_a}$ and $R_{s_b}$. To simplify matters we suppose that there is an exceptional state in the subsystem $A$ alone. Then according to (2.2) and (2.3) we shall have

$$R_{s_a}\!\left(\mathbf{x}_{s_a}^{(a)}, \mathbf{x}_{s_a}^{(a)'}\right) = \varphi_{s_a}\!\left(\mathbf{x}_{s_a}^{(a)}\right)\varphi_{s_a}^*\!\left(\mathbf{x}_{s_a}^{(a)'}\right) + \sum_{\nu_a} n_{s_a}\!\left(\varepsilon_{\nu_a}^{(s_a)}\right)\psi_{\nu_a}\!\left(\mathbf{x}_{s_a}^{(a)}\right)\psi_{\nu_a}^*\!\left(\mathbf{x}_{s_a}^{(a)'}\right),$$

$$R_{s_b}\!\left(\mathbf{x}_{s_b}^{(b)}, \mathbf{x}_{s_b}^{(b)'}\right) = \sum_{\nu_b} n_{s_b}\!\left(\varepsilon_{\nu_b}^{(s_b)}\right)\psi_{\nu_b}\!\left(\mathbf{x}_{s_b}^{(b)}\right)\psi_{\nu_b}^*\!\left(\mathbf{x}_{s_b}^{(b)'}\right). \qquad (2.11)$$

Upon substituting this into (2.10) and taking the formulae of Sec. 4 of I into account we shall obtain an expression of the type (2.4) only if Eq. (I.4.14) is fulfilled. For this reason we see that in the case considered in the present paper, too, the functions $n_s(z)$ are defined by (I.4.15)–(I.4.17) with two arbitrary parameters $\tau$ and $A$.

Now we turn to the relationship (I.4.1) as applied to $R_s^{(c)}$. Upon placing the expression for $R_s^{(c)}$ of (2.3) there and assuming that $s \ll N$ we obtain

$$\varphi_{s-1}(\mathbf{x}_{s-1})\,\varphi_{s-1}^*(\mathbf{x}_{s-1}') = \frac{1}{N}\int_V \varphi_s(\mathbf{x}_{s-1},\mathbf{r}_s)\,\varphi_s^*(\mathbf{x}_{s-1}',\mathbf{r}_s)d\mathbf{r}_s. \qquad (2.12)$$

Let us find $\varphi_1(\mathbf{r}_1)$ first. In the present paper, we restrict ourselves to spatially homogeneous boundless systems, in which case we can put $U_1 = 0$ (see (I.5.1)). Then (2.5) at $s = 1$ becomes

$$\frac{\hbar^2}{2m}\nabla^2\varphi_1(\mathbf{r}) + \varepsilon_{(1)}\varphi_1(\mathbf{r}) = 0. \qquad (2.13)$$

The solution of this equation, which gives $\rho_1^{(c)}$ = constant according to (2.6), is

$$\varphi_1(\mathbf{r}) = \sqrt{\rho_c} \exp\left(\frac{i}{\hbar}\mathbf{p}_0\mathbf{r}\right), \qquad \varepsilon_{(1)} = p_0^2/2m. \tag{2.14}$$

Here the quantity $\varepsilon_{(1)}$ is expressed in terms of another unknown quantity $p_0$, the magnitude of a vector $\mathbf{p}_0$; and the normalization factor is denoted by $\sqrt{\rho_c}$. We note for use later that a factor of the type $e^{i\alpha}$ plays no part for (2.3); for this reason, $\sqrt{\rho_c}$ can be considered to be a real number.

We turn next to Eq. (2.12) for the least possible value of $s$, namely $s = 2$, when $\mathbf{x}_{s-1}$ and $\mathbf{x}'_{s-1}$ are simply $\mathbf{r}_1$ and $\mathbf{r}'_1$ respectively. At given $\mathbf{r}_1$ and $\mathbf{r}'_1$, the main contribution to the integral in (2.12), as $V \to \infty$, results from regions where the absolute values $|\mathbf{r}_1 - \mathbf{r}_2|$ and $|\mathbf{r}'_1 - \mathbf{r}_2|$ are large (cf. the derivation of Eq. (I.4.8)). In this case, $U_2 \to 0$ according to (I.5.1), and Eq. (2.5) becomes

$$\frac{\hbar^2}{2m}\left(\nabla_1^2 + \nabla_2^2\right)\varphi_2^{(\infty)}(\mathbf{r}_1,\mathbf{r}_2) + \varepsilon_{(2)}\,\varphi_2^{(\infty)}(\mathbf{r}_1,\mathbf{r}_2) = 0, \tag{2.15}$$

the superscript $(\infty)$ denoting the limiting value of $\varphi_s$. The solution of (2.15) that meets the requirement of the symmetry can be taken to be

$$\varphi_2^{(\infty)}(\mathbf{r}_1,\mathbf{r}_2) = \frac{B_2}{2}\left\{\exp\left[\frac{i}{\hbar}(\mathbf{q}_1\mathbf{r}_1 + \mathbf{q}_2\mathbf{r}_2)\right] \pm \exp\left[\frac{i}{\hbar}(\mathbf{q}_1\mathbf{r}_2 + \mathbf{q}_2\mathbf{r}_1)\right]\right\}, \tag{2.16}$$

in which $B_2$ is presumed to be real (see above), while $\varepsilon_{(2)} = \left(q_1^2 + q_2^2\right)/2m$. Although in the present paper we deal solely with bosons, to make a comparison we have also written (2.16) for fermions to which the lower sign refers.

If (2.14) and (2.16) are put into (2.12) at $s = 2$, the following may be deduced immediately. Firstly, Eq. (2.12) can be satisfied only if $\mathbf{q}_1 = \mathbf{q}_2 = \mathbf{p}_0$ (then $\varepsilon_{(2)} = p_0^2/m$). Secondly, the right-hand side of (2.12) with the lower sign in (2.16) vanishes in the limit as $V \to \infty$ (and $N \to \infty$) because the integral does not increase proportionally to $V$ since $\varphi_2^{(\infty)} = 0$ in this case if $\mathbf{q}_1 = \mathbf{q}_2$. Hence, Eq. (2.12) entails $\rho_c = 0$. As should be expected, in the case of spinless fermions the condensate cannot be formed; and the fermions will not be considered furthermore. As to bosons, Eq. (2.12) yields $B_2 = \sqrt{\rho\rho_c}$ with $\rho = N/V$.

Investigating Eq. (2.12) for arbitrary $s$ we analyse first the case in which $|\mathbf{r}_j - \mathbf{r}_k|$ and $|\mathbf{r}'_j - \mathbf{r}'_k|$ are large if $j, k \le s - 1$ and $j \ne k$. Reasoning exactly as above, in calculation of the integral in (2.12) we can assume that $|\mathbf{r}_j - \mathbf{r}_s|$ and $|\mathbf{r}'_j - \mathbf{r}_s|$ are large as well (provided that the $|\mathbf{r}_j|$'s with

$j < s$ remain finite, though large). In this event, $U_s(\mathbf{x}_s) \to 0$ owing to (I.4.7) and (I.5.1); and the asymptotic symmetrized solution of (2.5) will be of the form

$$\varphi_s^{(\infty)}(\mathbf{x}_s) = \frac{B_s}{s!} \sum_P^{(s)} \exp\left(\frac{i}{\hbar} \sum_{k=1}^s \mathbf{r}_k \mathsf{P}\mathbf{q}_k\right). \tag{2.17}$$

We have seen that $\mathbf{q}_k = \mathbf{p}_0$ if $s = 1$ and 2. Substituting (2.17) into (2.12) shows that then $\mathbf{q}_k = \mathbf{p}_0$ for all $s$ and that $B_{s+1} = B_s \sqrt{\rho}$. Since $B_1 = \sqrt{\rho_c}$ and $B_2 = \sqrt{\rho \rho_c}$, one readily finds all the $B_s$'s. As a result of this, the asymptotic form of $\varphi_s$ and the quantities $\varepsilon_{(s)}$ are given by

$$\varphi_s^{(\infty)}(\mathbf{x}_s) = \sqrt{\rho_c}\, \rho^{(s-1)/2} \exp\left(\frac{i}{\hbar} \mathbf{p}_0 \sum_{k=1}^s \mathbf{r}_k\right), \qquad \varepsilon_{(s)} = \frac{s p_0^2}{2m}. \tag{2.18}$$

We turn now to the general case in which $\mathbf{r}_j$ and $\mathbf{r}'_j$ in (2.12) are arbitrary. The main contribution to the integral in (2.12) is again due to the limiting values of the functions as $|\mathbf{r}_s| \to \infty$. In this limit, from Eq. (I.4.7) it follows that $U_s(\mathbf{x}_s) \to U_{s-1}(\mathbf{x}_{s-1})$ (note that Eq. (I.4.7) is a consequence of (I.3.1), i.e. of (2.9)). Then, the argument $\mathbf{r}_s$ disappears from Eq. (2.5) and we can look for its solution in the form

$$\varphi_s(\mathbf{x}_s) = \tilde{\varphi}_{s-1}(\mathbf{x}_{s-1})\, \varphi_1(\mathbf{r}_s) = \sqrt{\rho_c}\, \tilde{\varphi}_{s-1}(\mathbf{x}_{s-1}) \exp\left(\frac{i}{\hbar} \mathbf{p}_0 \mathbf{r}_s\right), \tag{2.19}$$

where the last expression is written on account of (2.14). If this expression and the relation $\varepsilon_{(s)} - p_0^2/2m = \varepsilon_{(s-1)}$ that follows from (2.18) are substituted into (2.5), we shall see that the equation for $\tilde{\varphi}_{s-1}(\mathbf{x}_{s-1})$ coincides with the one for $\varphi_{s-1}(\mathbf{x}_{s-1})$, that is to say, $\tilde{\varphi}_{s-1}(\mathbf{x}_{s-1}) = C \varphi_{s-1}(\mathbf{x}_{s-1})$. The constant $C$ may be easily evaluated by use of (2.18), so that $C = \sqrt{\rho/\rho_c}$. Now, upon placing (2.19) in the right-hand side of (2.12), one verifies that it coincides with the left-hand side.

Summarizing, we see that Eq. (2.12) is satisfied for any $s$ if the solution of Eq. (2.5) is subject to the conditions of (2.18) from which the normalization of $\varphi_s(\mathbf{x}_s)$ follows as well. It is convenient to introduce $u_s(\mathbf{x}_s)$ instead of $\varphi_s(\mathbf{x}_s)$ according to

$$\varphi_s(\mathbf{x}_s) = \sqrt{\rho_c}\, \rho^{(s-1)/2}\, u_s(\mathbf{x}_s) \exp\left(\frac{i}{\hbar} \mathbf{p}_0 \sum_{k=1}^s \mathbf{r}_k\right). \tag{2.20}$$

It may be observed that this formula is valid at $s = 1$, too, with $u_1 = 1$, which is seen from (2.14). Eq. (2.5) yields the following equation for $u_s(\mathbf{x}_s)$

$$\frac{\hbar^2}{2m} \sum_{j=1}^s \nabla_j^2\, u_s(\mathbf{x}_s) + \frac{i\hbar \mathbf{p}_0}{m} \sum_{j=1}^s \nabla_j\, u_s(\mathbf{x}_s) - U_s(\mathbf{x}_s)\, u_s(\mathbf{x}_s) = 0. \tag{2.21}$$

By (2.18), the solution of this equation is to be subject to the condition that $u_s(\mathbf{x}_s) \to 1$ as the magnitude of all differences $\mathbf{r}_j - \mathbf{r}_k$ tends to infinity, that is, as $U_s(\mathbf{x}_s) \to 0$. Such a problem is typical of problems one meets in theory of particle scattering.

In homogeneous media we are dealing with, the vector $\mathbf{p}_0$ disappears in fact from (2.21). The reason for this is that, in such a medium, any physical quantity must not change if one makes the replacement $\mathbf{r}_k \to \mathbf{r}_k + \mathbf{a}$ with an arbitrary vector $\mathbf{a}$. Putting, for instance, $\mathbf{a} = -\mathbf{r}_1$ we obtain

$$u_s = u_s(0, \mathbf{r}_2 - \mathbf{r}_1, \mathbf{r}_3 - \mathbf{r}_1, \ldots, \mathbf{r}_s - \mathbf{r}_1), \quad U_s = U_s(0, \mathbf{r}_2 - \mathbf{r}_1, \mathbf{r}_3 - \mathbf{r}_1, \ldots, \mathbf{r}_s - \mathbf{r}_1), \quad (2.22)$$

so that Eq. (2.21) reduces to

$$\frac{\hbar^2}{m} \sum_{2 \leq j \leq k \leq s} \nabla_j \nabla_k \, u_s(\mathbf{x}_s) - U_s \, u_s(\mathbf{x}_s) = 0. \quad (2.23)$$

It will be noted that one can get rid of the mixed derivatives in this equation by using Jacobi's coordinates. Inasmuch as the limiting value of $u_s$ is real, solutions of (2.23) will be real as well. In what follows, only the case $s = 2$ will be needed for our aims. Then, Eq. (2.23) is similar to the Schrödinger equation for zero energy eigenvalue

$$\frac{\hbar^2}{m} \nabla^2 u_2(\mathbf{r}) - U_2(\mathbf{r}) \, u_2(\mathbf{r}) = 0, \qquad \mathbf{r} = \mathbf{r}_2 - \mathbf{r}_1, \quad (2.24)$$

with the condition that $u_2(\mathbf{r}) \to 1$ as $|\mathbf{r}| \to \infty$. This is equivalent to a problem such as the scattering of a particle with the propagation vector $\mathbf{k} = 0$ by a fixed force field. With the help of the Lippmann-Schwinger equation (see, e.g., [14]) or by using other methods, Eq. (2.24) together with the limiting condition can be reduced to an integral equation, the equation being

$$u_2(\mathbf{r}) + \frac{m}{4\pi\hbar^2} \int \frac{U_2(\mathbf{r}') \, u_2(\mathbf{r}')}{|\mathbf{r} - \mathbf{r}'|} d\mathbf{r}' = 1. \quad (2.25)$$

This last equation shows in particular that, as $|\mathbf{r}| \to \infty$, the function $u_2(\mathbf{r})$ differs from the limiting value by an amount proportional to $1/|\mathbf{r}|$, which conforms, as it should, to scattering theory. In this connection the noteworthy fact is the following. Into Eq. (2.12) were substituted the limiting values (2.17) and (2.19) in complete analogy with the derivation of Eq. (I.4.7) in Appendix C of I. In the case under consideration, however, the notion of the volume $V_0$ from Appendix C of I cannot be utilized because the functions $\varphi_s(\mathbf{x}_s)$ tend slowly to their limiting values. Nevertheless the use of the limiting values in (2.12) is fully legitimate for the integral of those values is proportional to the volume $V$ while the integral of terms in $|\mathbf{r}|^{-n}$ with $n < 3$ is proportional to $V^{1-n/3}$, so that the latter integral can be neglected in comparison with the former when $V \to \infty$, that is, in the thermodynamic limit.

Substituting (2.20) into (2.3) entails

$$R_s^{(c)}\left(\mathbf{x}_s, \mathbf{x}_s'\right) = \rho_c\, \rho^{s-1} u_s(\mathbf{x}_s)\, u_s(\mathbf{x}_s')\, \exp\left[\frac{i}{\hbar}\mathbf{p}_0 \sum_{k=1}^{s}\left(\mathbf{r}_k - \mathbf{r}_k'\right)\right]. \quad (2.26)$$

In Appendix B of I it has been proven that Eq. (I.2.14) at $s = 1$ is satisfied identically in the event of an uniform medium. One easily verifies that the same occurs if (2.26) at $s = 2$ is placed in (I.B.1) as $R_2$.

It is of interest to calculate $R_1^{(c)}$ by putting $s = 1$ in (2.26) and taking into account that $u_1 = 1$:

$$R_1^{(c)}(\mathbf{r}, \mathbf{r}') = \rho_c \exp\left[\frac{i}{\hbar}\mathbf{p}_0(\mathbf{r} - \mathbf{r}')\right]. \quad (2.27)$$

This expression shows that there is off-diagonal long range order (ODLRO, see Introduction) in $R_1(\mathbf{r},\mathbf{r}')$. If $\mathbf{p}_0 = 0$, the expression coincides with the relevant expression of [15] for an ideal Bose gas below the condensation point (note that our quantity $\rho_c$ corresponds to $\rho - \rho_c$ of [15]). Let us write down also the final formula for $\rho_s(\mathbf{x}_s)$ by inserting (2.20) into (2.7):

$$\rho_s(\mathbf{x}_s) = \rho_c \rho^{s-1} u_s^2(\mathbf{x}_s)$$
$$+ \frac{1}{2\pi i (2\pi\hbar)^{3s} s!} \int d\mathbf{m}_s \int_C dz\, n_s(z)\, v_s(\mathbf{x}_s,\mathbf{m}_s,z) \sum_P^{(s)} \exp\left[\frac{i}{\hbar}\sum_{k=1}^{s} \mathbf{r}_k(\mathbf{p}_k - \mathsf{P}\mathbf{p}_k)\right]. \quad (2.28)$$

Thus, Eqs. (2.7) and (2.5) can be replaced by (2.28) and (2.23). To these must be added (2.8) and (2.9) together with (I.4.15) and (I.4.17) determining $n_s(z)$. It should be emphasized that no approximation was used when deriving the hierarchy. The closed hierarchy so obtained contains four arbitrary constants $\tau$, $A$, $\rho_c$ and $\mathbf{p}_0$. At the same time, the vector $\mathbf{p}_0$ does not figure explicitly in the hierarchy but manifests itself in the off-diagonal elements of (2.26). In the case considered in I, such an hierarchy included two arbitrary constants $\tau$ and $A$ that were determined from thermodynamics and the normalization. It is natural to resort to an analogous procedure to determine $\tau$, $A$, $\rho_c$ and $\mathbf{p}_0$, which will be taken up in the next section.

Before closing this section let us discuss the physical bearing of the results obtained. To this end by making use of a quantum mechanical formula and (I.2.1), we calculate the momentum of the system (cf. (I.3.5) and (I.3.6))

$$\mathbf{P} = \int \Psi^*(\mathbf{x}_N)\left[-i\hbar \sum_{j=1}^{N}\nabla_j\right]\Psi(\mathbf{x}_N)\, d\mathbf{x}_N = -i\hbar \int \left[\nabla R_1(\mathbf{r},\mathbf{r}')\right]_{\mathbf{r}'=\mathbf{r}} d\mathbf{r}. \quad (2.29)$$

Upon inserting (2.27) we obtain that $\mathbf{P} = \rho_c V \mathbf{p}_0$ (the quantity $R_1^{(n)}$ does not contribute to (2.29), which can be seen from Eq. (3.3) below). Hence, the situation is analogous with that in which there are $N_c = \rho_c V$ particles that move at a speed of $\mathbf{p}_0/m$, although one cannot, of course, distinguish between the moving particles and the others owing to the principle of indistinguishability of identical particles. Such a state of the system is thermodynamically equilibrium according to the form of the density matrices as given by (2.1), and for this reason the existence of a flow in the fluid is not accompanied by any dissipative processes. Further aspects of this interpretation of superfluidity will be discussed in the concluding section.

### 3. The thermodynamics of a superfluid Bose-system

To begin with, we consider the above hierarchy at $s = 1$. Implying a spatially uniform medium we put $V^{(e)} = 0$; then Eq. (2.9) is satisfied identically while (2.8) gives (I.5.2). As long as $u_1 = 1$ and $\rho_1 = \rho \equiv N/V$ (this is a consequence of the normalization (I.2.3)), Eq. (2.28) reduces to

$$\rho = \rho_c + \frac{\sqrt{\pi}}{2} A\omega\tau^{3/2}, \tag{3.1}$$

where we have used $n_1(z)$ of (I.4.16). Here and henceforth $\omega$ denotes the same constant as in (I.5.4). It is convenient to employ the quantity $\rho_n = \rho - \rho_c$ instead of $A$ because, on account of (3.1),

$$\rho_n = \frac{\sqrt{\pi}}{2} A\omega\tau^{3/2}. \tag{3.2}$$

By (I.5.2), Eq. (I.2.20) enables one to calculate $R_1^{(n)}(\mathbf{r},\mathbf{r}')$, which combined with (2.27) yields

$$R_1(\mathbf{r},\mathbf{r}') = \rho_c \exp\left[\frac{i}{\hbar}\mathbf{p}_0(\mathbf{r}-\mathbf{r}')\right] + \rho_n \exp\left[-\frac{m\tau(\mathbf{r}-\mathbf{r}')^2}{2\hbar^2}\right]. \tag{3.3}$$

If $\mathbf{p}_0 \neq 0$, the system cannot be isotropic. When considering the pair correlation function $g(\mathbf{r}) = \rho_2(\mathbf{r})/\rho^2$ where $\mathbf{r}$ is the same as in (2.24), we shall assume a cylindrical symmetry. We orient the $z$ axis in the direction specified by $\mathbf{p}_0$ and denote the distance perpendicular to the $z$ axis by $\rho_{xy} = (x^2 + y^2)^{1/2}$. Then, $g(\mathbf{r})$ will have a spatial dependence of the form $g = g(\rho_{xy},z)$. It will be noted that $g(\mathbf{r}) = g(-\mathbf{r})$ by virtue of the relation $\rho_2(\mathbf{r}_1,\mathbf{r}_2) = \rho_2(\mathbf{r}_2,\mathbf{r}_1)$. For what follows we introduce the two functions

$$\bar{g}(r) = \frac{1}{r}\int_0^r g\left(\sqrt{r^2-z^2},z\right)dz, \quad \tilde{g}(r) = \frac{3}{r^3}\int_0^r g\left(\sqrt{r^2-z^2},z\right)z^2\,dz \tag{3.4}$$

with $r = |\mathbf{r}|$. It should be remarked that in the case of a spherical symmetry

$$\bar{g}(r) = \tilde{g}(r) = g(r). \tag{3.5}$$

The internal energy $E$ is given by (I.3.6). Upon inserting (3.3) one obtains, instead of (I.5.16),

$$E = \frac{p_0^2}{2m}\rho_c V + \tfrac{3}{2}\tau\rho_n V + 2\pi N\rho \int_0^\infty r^2 K(r)\overline{g}(r)\,dr. \tag{3.6}$$

Formally, this expression corresponds to a situation in which $N_n = \rho_n V$ particles take part in the thermal motion while the remaining $N_c = \rho_c V$ particles travel with the velocity $p_0/m$. Physically, all the particles participate in the thermal motion because of the interaction and because of the quantum effects mentioned when discussing (2.29).

Inasmuch as the system is anisotropic, one must employ the stress tensor instead of the pressure. Derivation of a general formula for the stress tensor with account taken of the fact that the exact wavefunction $\Psi(\mathbf{x}_N,t)$ depends upon the time (see I) is given in Appendix A. For a homogeneous whilst anisotropic medium, Eq. (A.10) yields

$$\sigma_{ij} = \frac{\hbar^2}{mV}\int\left[\frac{\partial^2}{\partial x_i \partial x_j}R_1(\mathbf{r},\mathbf{r}')\right]_{\mathbf{r}'=\mathbf{r}} d\mathbf{r} + \frac{1}{2}\int x_j \frac{\partial K(|\mathbf{r}|)}{\partial x_i}\rho_2(\mathbf{r})\,d\mathbf{r}. \tag{3.7}$$

It is worthy of remark that, in the case of an isotropic medium when $\sigma_{ij} = -p\delta_{ij}$, Eq. (3.7) goes over into (I.3.7). Let $p = -\sigma_{xx} = -\sigma_{yy}$ be the pressure on a surface element normal to the $x$ or $y$ axis, and $p + \Delta p = -\sigma_{zz}$ the one on such an element normal to the $z$ axis. By analogy with (3.6), from (3.7) it follows that

$$p = \rho\tilde{\tau} - \frac{\pi\rho^2}{3}\int_0^\infty r^3 \frac{dK}{dr}\left[3\overline{g}(r) - \tilde{g}(r)\right]dr, \tag{3.8}$$

$$\Delta p = \rho_c \frac{p_0^2}{m} - \pi\rho^2 \int_0^\infty r^3 \frac{dK}{dr}\left[\tilde{g}(r) - \overline{g}(r)\right]dr, \tag{3.9}$$

where the notation $\tilde{\tau} = \rho_n\tau/\rho$ is used. Hereafter, in place of $\rho_c$ and $\rho_n$ it is more convenient to introduce $f_c = \rho_c/\rho = 1 - \rho_n/\rho$ so that

$$\tilde{\tau} = (1 - f_c)\tau. \tag{3.10}$$

According to the ideas of I, after these preliminary steps we turn to the second law of thermodynamics for quasi-static processes, which will permit us to construct thermodynamics. First of all, we need the expression for the work with account taken of the fact that the system is not isotropic as there is a preferred direction specified by $\mathbf{p}_0$. There is no loss of generality in supposing that the system is enclosed in a cylinder whose axis is directed along $\mathbf{p}_0$, the length of the cylinder being $L$ and its cross-section $V/L$. The work done in changing the dimensions of the cylinder is

$$\delta A = p\,dV + \Delta p \frac{V}{L}\,dL. \tag{3.11}$$

Now the second law of thermodynamics ($\theta dS = dE + \delta A$ [16]) assumes the form

$$\theta dS = \frac{\partial E}{\partial \theta} d\theta + \left(\frac{\partial E}{\partial V} + p\right) dV + \left(\frac{\partial E}{\partial L} + \Delta p \frac{V}{L}\right) dL. \tag{3.12}$$

The condition that the entropy $S$ shall exist in accordance with (3.12), i.e., the equality of the corresponding second derivatives, yields now two more equations in addition to Eq. (I.5.19). As a result, we have the following three equations

$$\frac{\rho^2}{N}\left(\frac{\partial E}{\partial \rho}\right)_\theta = p - \left(\frac{\partial p}{\partial \theta}\right)_\rho, \tag{3.13}$$

$$\theta \frac{\partial \Delta p}{\partial \theta} - \Delta p = 0, \qquad \rho \frac{\partial \Delta p}{\partial \rho} - \Delta p = 0. \tag{3.14}$$

Here we have taken account of the fact that, at a given $V$, neither $E$ nor $p$ depend on $L$ according to (3.6) and (3.8).

The two equations of (3.14) have a unique simultaneous solution $\Delta p = C_0 \theta \rho$ where $C_0$ is a constant independent of $\theta$ and $\rho$, and thereby of $\mathbf{p}_0$. If $\mathbf{p}_0 = 0$, the system is isotropic and one should have $\Delta p = 0$, which shows that $C_0 = 0$. Consequently, always $\Delta p = 0$ even if $\mathbf{p}_0 \neq 0$, that is to say, notwithstanding the existence of a flow in the superfluid, Pascal's law is fulfilled and all thermodynamic relations, e.g., Eq. (3.11), are of the usual form. Now from Eq. (3.9) it follows that

$$\rho_c\, p_0^2 = \pi m \rho^2 \int_0^\infty r^3 \frac{dK}{dr} \left[\tilde{g}(r) - \bar{g}(r)\right] dr. \tag{3.15}$$

By virtue of this the expression for $E$, Eq. (3.6), transforms to

$$E = \tfrac{3}{2}\tilde\tau N + 2\pi N\rho \int_0^\infty r^2 K(r)\bar{g}(r)\, dr + \frac{\pi}{2} N\rho \int_0^\infty r^3 \frac{dK}{dr}\left[\tilde{g}(r) - \bar{g}(r)\right] dr. \tag{3.16}$$

Upon substituting Eqs. (3.8) and (3.16) into Eq. (3.13) and integrating by parts, one obtains the equation

$$2\theta \frac{\partial \tilde\tau}{\partial \theta} + 3\rho \frac{\partial \tilde\tau}{\partial \rho} - 2\tilde\tau = \frac{\pi\rho}{3} \int_0^\infty dr\, r^2 \left\{ K(r)\left[3(\tilde{g}-\bar{g}) - 12\rho \frac{\partial \bar{g}}{\partial \rho} + r\left(3\frac{\partial \bar{g}}{\partial r} + \frac{\partial \tilde{g}}{\partial r}\right)\right]\right.$$

$$\left. + r\frac{dK}{dr}\left[2\theta\left(3\frac{\partial \bar{g}}{\partial \theta} - \frac{\partial \tilde{g}}{\partial \theta}\right) + 3\rho\left(\frac{\partial \bar{g}}{\partial \rho} - \frac{\partial \tilde{g}}{\partial \rho}\right)\right]\right\}. \tag{3.17}$$

If this last equation is solved, the quantity $p_0$ can be calculated with the help of (3.15), the quantity $f_c$ being found from (3.10). It is worth noting that Eq. (3.17) passes into Eq. (I.5.20) if $f_c = 0$ since then $\tilde\tau = \tau$ by (3.10) and one has (3.5).

Before discussing Eq. (3.17) let us consider the Helmholtz free energy $F$. In Appendix B it is shown that $F$ can be calculated by

$$F(\theta,\rho) = -\theta \int_{\theta_0}^{\theta} E(x,\rho)\frac{dx}{x^2} + \frac{\theta}{\theta_0} N \int_{\rho_0}^{\rho} p(\theta_0,y)\frac{dy}{y^2} + \frac{\theta}{\theta_0} F(\theta_0,\rho_0), \qquad (3.18)$$

where $\theta_0$ and $\rho_0$ are arbitrary constants. In our case, $E$ and $p$ are known by virtue of (3.16) and (3.8). In Appendix B it is shown as well that one may also use the expression

$$F(\theta,\rho) = 2\pi\rho N \int_0^\infty dr\, r^2 \left(2K + r\frac{dK}{dr}\right) \int_1^{\sqrt{\theta_0/\theta}} d\xi\, \bar{g}(r,\theta\xi^2,\rho\xi^3) + \frac{\theta}{\theta_0} N \int_{\rho_0}^{\rho(\theta_0/\theta)^{3/2}} p(\theta_0,y)\frac{dy}{y^2}$$

$$+ \frac{\theta}{\theta_0} F(\theta_0,\rho_0). \qquad (3.19)$$

Now we discuss Eq. (3.17). This equation is, in fact, an equation for two unknown functions $\tilde{\tau}(\theta,\rho)$ and $\tau(\theta,\rho)$ because of $\tau$-dependence of $g(r)$ via $n_2(\tau)$. In what follows it is more convenient to work with $\tilde{\tau}(\theta,\rho)$ and $f_c(\theta,\rho)$, instead of $\tilde{\tau}$ and $\tau$, upon eliminating $\tau$ by (3.10). In addition to Eq. (3.17), it is necessary to obtain a second equation relating $\tilde{\tau}$ and $f_c$. To do this, one can proceed from the fact that at given $\theta$ and $\rho$ the state of thermal equilibrium corresponds to a minimum of the Helmholtz free energy $F$. In this question it is much more convenient to use Eq. (3.19) for $F$ rather than (3.18) since, if one follows the curve $\rho = \rho_0(\theta/\theta_0)^{3/2}$ in the $\rho$–$\theta$ diagram, then the second integral of (3.19) vanishes and all essential variation of $F$ will be due to the first term because the last term, linear in $\theta$, is insignificant for thermodynamics. At the same time it should be observed that, when considering variations of $F$ connected with those of $f_c$, it must be kept in mind that $F(\theta_0,\rho_0)$, too, may depend on $f_c(\theta_0,\rho_0)$. To simplify matters one may assume $\theta_0$ and $\rho_0$ to be such that $f_c(\theta_0,\rho_0) = 0$, in which case $F(\theta_0,\rho_0)$ will not contain $f_c$. As to (3.18), that formula will be rather complicated after substituting (3.8) and (3.16).

According to (2.28) (see also (5.1) below) the quantity $\bar{g}$ is a functional of $\tilde{\tau}$ and $f_c$ (it depends on $\rho$ as well but the temperature $\theta$ does not enter into (2.28) explicitly), that is to say, $\bar{g} = \bar{g}(r,\tilde{\tau},f_c,\rho)$. Note that $\bar{g}$ contains no derivatives of $\tilde{\tau}$ and $f_c$. It also worth remarking that Eq. (3.17) should not be considered to be an accessory constraint under which one seeks a minimum of $F$ because the integration in (3.19) is carried out along a line in the $\rho$–$\theta$ phase plane while any relation between $\tilde{\tau}$ and $f_c$ on a line may play the role of a boundary condition for (3.17) at most. We change $f_c$ and consider the corresponding variations of $F$. Eq. (3.19), with account taken of the foregoing, yields

$$\delta F = 2\pi\rho N \int_1^{\sqrt{\theta_0/\theta}} d\xi\, \delta f_c(\theta\xi^2,\rho\xi^3)\frac{\partial}{\partial f_c}\int_0^\infty dr\, r^2\left(2K + r\frac{dK}{dr}\right)\bar{g}(r,\tilde{\tau},f_c,\rho\xi^3) \qquad (3.20)$$

with $\tilde{\tau} = \tilde{\tau}(\theta\xi^2,\rho\xi^3)$ and $f_c = f_c(\theta\xi^2,\rho\xi^3)$.

At high temperatures, the coefficient of $\delta f_c$ in (3.20) has to be positive, naturally, providing a minimum of $F$ at $f_c = 0$. Eq. (3.17) in this case transforms to (I.5.20) as mentioned below (3.17), and we arrive at the formulae of I that describe the high-temperature phase. At the point of the supposed phase transition the coefficient must vanish. Below the transition point, when $f_c \neq 0$, the condition of minimum $F$, that is, $\delta F = 0$ in (3.20), entails

$$\frac{\partial}{\partial f_c} \int_0^\infty r^2 \left( 2K + r \frac{dK}{dr} \right) \bar{g}(r,\tilde{\tau},f_c,\rho) dr = 0, \tag{3.21}$$

where $\tilde{\tau} = \tilde{\tau}(\theta,\rho)$ and $f_c = f_c(\theta,\rho)$ inasmuch as $\theta\xi^2$ and $\rho\xi^3$ are arbitrary for arbitrary $\theta$ and $\rho$. If one introduces the relation between $\tilde{\tau}$ and $f_c$ as given by (3.21), together with (3.10), into (3.17), one will obtain an equation that will contain only one unknown function $\tilde{\tau}(\theta,\rho)$.

The transition point, more precisely the line of transition points in the $\rho$–$\theta$ plane, is given by (3.21) provided that one puts $f_c = 0$, $\bar{g} = g$ (see (3.5)) and $\tilde{\tau} = \tau(\theta,\rho)$ with $\tau(\theta,\rho)$ relevant to the high-temperature phase. The boundary condition for the equation resulting from (3.17) after substituting $f_c$ as given by (3.21) is the equality $\tilde{\tau} = \tau$ on the line obtained. From (3.10) it follows automatically that $f_c = 0$ on the line.

When comparing the above argumentation with the one used in the Landau phenomenological theory of phase transitions [17], one sees that the argumentations are akin to each another differing in that $f_c > 0$ in our case, and thereby $F$ as a functional of $f_c$ can contain a linear term. The order parameter in the Landau theory can be of any sign, so that the expansion of a thermodynamic potential in powers of the order parameter begins at quadratic terms. In other words, one can say that the role of the order parameter in the present theory is played by $\pm\sqrt{f_c}$.

From the mathematical point of view, the problem under consideration amounts to a problem of searching for a solution of nonlinear equations that branches off from a known solution. The line that was called the phase transition line to simplify reasoning a bit, ought to be termed the line of bifurcation points because the continuity of thermodynamic quantities on that line need not signify that one will really observe a second-order phase transition and this will occur just on the line in question. To find out the order of the phase transition and obtain the actual transition line it is yet necessary to investigate isotherms of the low-temperature phase. All the foregoing general arguments will be illustrated by concrete examples in Secs. 6 and 7.

Concluding the section we observe that the thermodynamics of a superfluid Bose system does in no way differ from the customary equilibrium thermodynamics. For such a system, all thermodynamic relations remain valid in spite of the existence of a superflow. Even Pascal's law is fulfilled.

## 4. Stability limit

In the problem under discussion another situation is possible. Before the free energy reaches a minimum, there may happen a breakdown of the condition of the mechanical stability of the system, the condition being $(\partial p/\partial V)_\theta \leq 0$. Just this situation occurs in an ideal Bose gas. In order to demonstrate this let us consider first the free energy $F_h$ of the ideal Bose gas in the high-temperature phase, that is, above the condensation point. According to (I.6.23) at $n = 0$ and (I.5.25) (see also [16], [17]), $F_h(\theta,\rho)$ is given in a parametric form by

$$F_h = -N\theta \left[ \beta^2 + \frac{2G_1(\beta)}{3G_0(\beta)} \right], \qquad \rho = \omega\theta^{3/2} G_0(\beta), \qquad (4.1)$$

where we have put $\alpha = -\beta^2$. A plot of $F_h(\theta,\rho)$ as a function of $\theta$ is presented in figure 1 by curve 1. It should be stressed that at $\theta = \theta_i$ the function $F_h(\theta,\rho)$ is smooth and has no singularities according to Appendices F and G of I. In the low-temperature condensed phase, the free energy in the notation of I is, [16,17],

$$F_l = -\frac{2\omega}{3\rho} N \zeta_5 \theta^{5/2}. \qquad (4.2)$$

This dependence is represented in figure 1 by curve 2 (the curve makes no sense at $\theta > \theta_i$ since then $f_c < 0$). From figure 1 it follows that always $F_h \leq F_l$. For this reason it would seem that the condensed phase should never be observed. However, it is easy to show that on branch 1 we have $(\partial p/\partial V)_\theta > 0$ at $\theta < \theta_i$, and the relevant phase in which $f_c = 0$ is unstable. Hence, at $\theta < \theta_i$ is realized the condensed phase that does not correspond to a minimum of $F$ but for which $(\partial p/\partial V)_\theta = 0$.

In the event of a nonideal Bose system, a similar situation is not excluded too as the example treated in Sec. 6 below shows. Inasmuch as we shall need relevant formulae in Sec. 6 and also having in mind that one may encounter such a situation in other instances, let us consider the general case of a system whose thermodynamics is governed not by the principle of minimum free energy but by the condition $(\partial p/\partial V)_\theta = 0$, that is,

$$\left( \frac{\partial p}{\partial \rho} \right)_\theta = 0. \qquad (4.3)$$

It should be emphasized that we imply a single-phase system, but not a two-phase system in a first-order phase transition when Eq. (4.3) is fulfilled as well.

Eq. (4.3) together with (3.13) constitute a set of two simultaneous equations for two functions $E(\theta,\rho)$ and $p(\theta,\rho)$. The set can be readily solved, and the general solution can be written as

$$E = N\frac{\theta^2}{\rho}\varphi_1'(\theta) + N\varphi_2(\theta), \qquad p = \theta\varphi_1(\theta), \tag{4.4}$$

where $\varphi_1(\theta)$ and $\varphi_2(\theta)$ are arbitrary functions of $\theta$, and $\varphi_1'(\theta) = d\varphi_1/d\theta$. Eq. (3.18) enables one to evaluate the free energy as well.

The functions $\varphi_1(\theta)$ and $\varphi_2(\theta)$ are uniquely determined provided one knows the internal energy $E_h(\theta,\rho)$ and the pressure $p_h(\theta,\rho)$ in the high-temperature phase. As the temperature is lowered, the high-temperature phase loses the stability when the condition $(\partial p/\partial V)_\theta < 0$ breaks down. Let $\theta_1$ and $\rho_1$ be the corresponding temperature and density (we follow the notation of I; for an ideal Bose gas $\theta_1 = \theta_i$). These last values are found from Eq. (4.3) written for the high-temperature phase and yield an equation of the phase transition line in the $\theta$–$\rho$ plane[1], say

$$\theta_1 = \theta_1(\rho_1) \quad \text{or} \quad \rho_1 = \rho_1(\theta_1). \tag{4.5}$$

On this curve, $E$ and $p$ should be continuous and coincide with the values given by (4.4), so that

$$E_h[\theta_1,\rho_1(\theta_1)] = N\frac{\theta_1^2}{\rho_1(\theta_1)}\varphi_1'(\theta_1) + N\varphi_2(\theta_1), \qquad p_h[\theta_1,\rho_1(\theta_1)] = \theta_1\varphi_1(\theta_1). \tag{4.6}$$

From these equations, we find the functions $\varphi_1(\theta)$ and $\varphi_2(\theta)$:

$$\varphi_1(\theta) = \frac{1}{\theta}p_h[\theta,\rho_1(\theta)], \qquad \varphi_2(\theta) = \frac{1}{N}E_h[\theta,\rho_1(\theta)] - \frac{\theta^2}{\rho_1(\theta)}\varphi_1'(\theta). \tag{4.7}$$

Now, by (4.4) and (3.18), we determine $E$, $p$ and $F$ in the low-temperature phase, which permits us to compute all its thermodynamic quantities. In the above reasoning, we have not referred to (3.8) and (3.16) (in the present instance Eq. (3.17) need not be considered for we used Eq. (3.13) equivalent to (3.17)). As long as we already know $E$ and $p$, Eqs. (3.8) and (3.16) enable us to calculate $\tilde{\tau}$ and $\tau$, and after this $f_c$ by (3.10) and $p_0$ by (3.15). The fact that $E$ and $p$ as given by (3.8) and (3.16) depend upon two parameters $\tilde{\tau}$ and $\tau$ proves to be essential here. If they depended upon only one parameter as in I, Eqs. (3.8) and (3.16) could be incompatible with one another.

Let us consider peculiarities of the phase transition as a result of which there emerges a phase characterized by (4.3). In the $p$–$\rho$–$\theta$ space, two surfaces $p = p(\theta,\rho)$ and $p = p_h(\theta,\rho)$ have a common curve given by (4.5), and Eq. (4.3) is satisfied on the curve for both the surfaces. Therefore, the surfaces touch each other along this curve, which amounts to saying that the derivative $(\partial p/\partial\theta)_\rho$ is the same on the curve for both the surfaces, that is to say, the derivative is continuous at $\theta = \theta_1$. Now from (3.13) which holds for both the phases, it follows that in the $E$–$\rho$–$\theta$ space the two surfaces $E = E(\theta,\rho)$ and $E = E_h(\theta,\rho)$ have coincident derivatives $(\partial E/\partial\rho)_\theta$ on the common curve (4.5) and thereupon the surfaces touch each other. Reasoning as above we see that $(\partial E/\partial\theta)_\rho$ is also continuous at $\theta = \theta_1$. As a result, we conclude that the isochoric heat capacity $C_V = (\partial E/\partial\theta)_\rho$ is continuous in the phase transition under discussion (its derivative may be discontinuous). The isobaric heat capacity $C_p$ can be calculated with the aid of the formula [17]

$$C_p = C_V - \theta \left(\frac{\partial p}{\partial \theta}\right)_V^2 \bigg/ \left(\frac{\partial p}{\partial V}\right)_\theta . \tag{4.8}$$

---

[1] In fact, this may be a line of bifurcation points as mentioned at the end of Sec. 3. In an ideal Bose gas, this line is actually the phase transition line.

Inasmuch as all the quantities on the right are finite on the transition line while $(\partial p/\partial V)_\theta \to 0$, the heat capacity $C_p$ of the high-temperature phase will diverge when approaching the transition point. This behaviour of $C_V$ and $C_p$ is analogous with that which occurs in the Bose condensation of an ideal gas [18].

If one applies (4.7) to an ideal Bose gas, one will get

$$\varphi_1(\theta) = \tfrac{2}{3}\zeta_5\omega\theta^{3/2}, \qquad \varphi_2(\theta) = 0. \tag{4.9}$$

When substituted into (4.4) these lead to the known expressions for $E$ and $p$ of a condensed ideal boson gas [16,17]. Eqs. (3.18) or (3.19) with $K \equiv 0$ yield (4.2) for $F$ as long as the point $(\theta_0,\rho_0)$ is taken to be a point on the curve (4.5) whose equation is $\rho_1 = \zeta_3\,\omega\,\theta_1^{3/2}$ in the present case. Hence, all thermodynamic properties of the condensed ideal gas may be obtained without using the Bose distribution for the condensed phase. However, we shall not find $f_c$ in this manner because $f_c$ figures in no thermodynamic formula, disappearing from (3.8), (3.16) and (3.19) when $K \equiv 0$.

It is worth remarking that, in the low-temperature phase under discussion, the pressure is independent of the volume in view of (4.3). For this reason the pressure will not increase even if the volume decreases down to zero. The volume of a real system cannot diminish without limit. This indicates that at high densities there must necessarily appear a minimum of the free energy in the stability region $(\partial p/\partial V)_\theta < 0$, so that the subsequent properties of the system will correspond to the free energy being a minimum.

## 5. Pair correlation function

The foregoing shows that of special interest is the pair correlation function $g(\mathbf{r}) = \rho_2(\mathbf{r})/\rho^2$ for it specifies all thermodynamic properties of an equilibrium system. By analogy with (I.5.7), Eq. (2.28) at $s = 2$ gives the following expression for this function

$$g(\mathbf{r}) = \frac{\rho_c}{\rho} u_2^2(\mathbf{r}) + \frac{1}{2\pi i (2\sqrt{2}\pi\hbar)^3 \rho} \int d\mathbf{q} \int_C ds\, n(s) v(\mathbf{r},\mathbf{q},s) \left[1 + \exp\left(\frac{i}{\hbar}\mathbf{q}\mathbf{r}\right)\right]. \quad (5.1)$$

The equation for $v(\mathbf{r},\mathbf{q},s)$ can be obtained from (2.8) in the same manner as (I.5.6), with the result

$$\frac{\hbar^2}{m}\nabla^2 v + \frac{i\hbar}{m}\mathbf{q}\nabla v + \left[s - \frac{\mathbf{q}^2}{4m} - U_2(\mathbf{r})\right] v = 1. \quad (5.2)$$

The function $u_2(\mathbf{r})$ that figures in (5.1) is determined by (2.24) or (2.25).

We note in passing that owing to the first term in (5.1) the decay of spatial correlations is slow as $|\mathbf{r}| \to \infty$: they decay proportionally to $1/|\mathbf{r}|$, which stems from the remark following (2.25). Therefore, the relevant state of the system is characterized by long-range correlations.

The potential $U_2(\mathbf{r})$ is specified by Eq. (2.9) at $s = 2$. In case the system is not isotropic, Eq. (I.5.33) no longer holds and, besides, a problem arises as to the integrability of equations like (2.9) [19]. The approximation for $\rho_3(\mathbf{r}_1,\mathbf{r}_2,\mathbf{r}_3)$ must be chosen such that the equation for $U_2(\mathbf{r})$ be integrable. It is preferable to utilize an approximation which leads directly to an equation for $U_2(\mathbf{r})$ because then the problem of integrability does not emerge. For example, an equation of the type (I.5.36) extended to an anisotropic medium may be used. If one takes some equivalent of Eq. (2.9) for $U_2(\mathbf{r})$ or other, one will obtain four equations for the four functions $g$, $u_2$, $v$ and $U_2$.

By analogy with Appendix D of I, we introduce a function $w(\mathbf{r},\mathbf{q},s)$ according to

$$v(\mathbf{r},\mathbf{q},s) = w(\mathbf{r},\mathbf{q},s) \exp\left(-\frac{i}{2\hbar}\mathbf{q}\mathbf{r}\right). \quad (5.3)$$

The equation for $w(\mathbf{r},\mathbf{q},s)$ follows from (5.2):

$$\frac{\hbar^2}{m}\nabla^2 w(\mathbf{r},\mathbf{q},s) + [s - U_2(\mathbf{r})] w(\mathbf{r},\mathbf{q},s) = \exp\left(\frac{i}{2\hbar}\mathbf{q}\mathbf{r}\right). \quad (5.4)$$

We look for a solution of this last equation in terms of an expansion in eigenfunctions $\psi_\nu(\mathbf{r})$ defined by

$$\frac{\hbar^2}{m}\nabla^2 \psi_\nu(\mathbf{r}) + [\varepsilon_\nu - U_2(\mathbf{r})] \psi_\nu(\mathbf{r}) = 0. \quad (5.5)$$

Proceeding in the same fashion as in Appendix D of I, we arrive at (I.D.6) for the second term in (5.1), so that

$$g(\mathbf{r}) = \frac{1}{\rho}\left[\rho_c\, u_2^2(\mathbf{r}) + 4\sqrt{2}\sum_\nu {}' n(\varepsilon_\nu)|\psi_\nu(\mathbf{r})|^2\right], \tag{5.6}$$

where the primed summation is to be extended over even functions $\psi_\nu(\mathbf{r})$. In the case of a continuous spectrum one must integrate.

Upon comparing (5.5) at $\varepsilon_\nu = \varepsilon_0 = 0$ with (2.24) we see that $u_2(\mathbf{r}) \propto \psi_0(\mathbf{r})$ (the normalizations of these functions are different). Consequently, instead of two Eqs. (2.24) and (5.2) we have only one Eq. (5.5). In fact, (5.5) is an infinite set of simultaneous nonlinear equations, each of them containing all $\psi_\nu(\mathbf{r})$'s and $\varepsilon_\nu$'s since $g(\mathbf{r})$ of (5.6) and thereby $U_2(\mathbf{r})$ are functionals of all $\psi_\nu$'s. In different cases, one may use either two equations (2.24) and (5.2) or the infinite set given by (5.5).

To obviate any misunderstanding it should be remarked the following. The pair correlation function $g(\mathbf{r})$ describes a relative distribution of two particles since $\mathbf{r} = \mathbf{r}_2 - \mathbf{r}_1$ in view of (2.24). If the particles move at the same speed (equal to $\mathbf{p}_0/m$ in our case), the exceptional level for this distribution is the level $\varepsilon_\nu = 0$ rather than $\varepsilon_{(2)} = p_0^2/m$ that figures in Eq. (2.15). This is reflected in Eq. (5.5) in which it is just the level $\varepsilon_\nu = 0$ that corresponds to $u_2(\mathbf{r})$.

The equations considered in the present section can have spherically symmetric solutions. Then $\mathbf{p}_0 = 0$ according to (3.5) and (3.15). Such a state of the system can be called the condensate phase without superfluidity[2]. Therefore, formation of a condensate does not necessarily lead to superfluidity. In particular, superfluidity cannot exist in an ideal gas because Eq. (3.15) gives $\mathbf{p}_0 = 0$ if $K \equiv 0$, on which point the present theory is in agreement with Landau's argument [2,3]. In Appendix C it is shown that superfluidity does not occur in a slightly nonideal Bose system as well. Consequently, superfluidity can be observed only if the interaction between atoms is sufficiently strong, in which case the equations obtained may have solutions without spherical symmetry. To find such solutions of a set that comprises both differential and integral equations is not a simple problem in which one cannot resort to traditional perturbation expansions as shown in Appendix C.

Let us present an argument that nonsymmetric solutions may exist. To this end we turn to Eq. (2.24) that coincides with the Schrödinger equation for zero energy eigenvalue. Even if the potential is spherically symmetric, besides spherically symmetric solutions the Schrödinger equation can, for special cases of the potential, have zero-energy solutions that are not spherically symmetric (see, e.g., Ref. [20, Sec. 15] where Schiff considers bound states that appear with zero energy; in scattering theory this corresponds to resonance scattering). Of course, an eigenvalue that is exactly zero can occur only in an exceptional case of the potential. However, our potential $U_2(\mathbf{r})$ is not a given quantity, and $U_2(\mathbf{r})$ may adjust itself to the required form. The adjustment can be facilitated if the interaction potential $K(\mathbf{r})$ has a real or virtual energy level near zero. Just such a level exists in the case of helium atoms [21,22]. Insofar as in nature there is only one Bose liquid that manifests superfluidity, experiment cannot provide an answer to the question whether superfluidity is a unique property of helium, which is completely due to peculiarities of the interaction between its atoms, or the class of potentials $K(\mathbf{r})$ allowing superfluidity is rather wide. The answer may be obtained by investigating the above hierarchy of equations.

---

[2] We shall employ the term "condensate phase" rather than the more familiar "condensed phase" in order to emphasize the existence of a condensate in the system.

On lowering the temperature there may appear various, both spherically symmetric and nonsymmetric, solutions of Eq. (2.24) which correspond to different values of the quantum number $l$, that is to say, of the angular momentum in the usual interpretation of the Schrödinger equation (it should be noted that $\mathbf{p}_0 \neq 0$ only if $l \neq 0$). Consequently, condensate phases of different types are possible, both superfluid and without superfluidity. Clearly, there will occur the phase that provides an absolute minimum of the appropriate thermodynamic potential. The situation is similar to that which is met with in theory of a crystal where, for a given interaction potential, various crystalline lattices are possible while only one of them is realized.

## 6. Hard spheres under the neglect of triplet correlations

In I, two examples of solution of the equations obtained were considered. In the case of a Bose system, those solutions describe high-temperature phases and do not exist as $\theta \to 0$. It is natural to resort to the same examples in the present paper as well in order to illustrate the general results obtained above and to find out properties of the pertinent low-temperature phases together with peculiarities of the phase transitions involved.

In the present section, we shall deal with a hard-sphere system wherein triplet correlations are neglected, which case was treated in Sec. 6 of I. In this context it should be mentioned that, for a hard-sphere potential, the method of $\lambda$ expansions considered in Appendix C is of no use because the potential is extremely singular (formally, $\lambda = \infty$). If $U_2(\mathbf{r}) = K(\mathbf{r})$ with $K(\mathbf{r})$ defined by (I.6.1), Eq. (2.24) has a unique solution of the form

$$u_2(r) = 1 - \frac{a}{r}, \quad r = |\mathbf{r}| \tag{6.1}$$

subject to two conditions, namely, $u_2(a) = 0$ and $u_2(r) \to 1$ as $r \to \infty$.

In the approximation used, $U_2(\mathbf{r})$ does not depend on $u_2(\mathbf{r})$, so that Eq. (5.2) and its solution will be the same as in I. Therefore, in the case under study the second term of (5.1) will differ from (I.6.3) only because of the value of $A$ given now by (3.1). Upon denoting the right-hand side of (I.6.3) with $A$ from (I.5.3) by $g_I(r)$ we shall obtain, as a result,

$$g(r) = 1 - \frac{2a}{r} f_c \left(1 - \frac{a}{2r}\right) + (1 - f_c)\left[g_I(r) - 1\right]. \tag{6.2}$$

Since $g_I(r) \to 1$ as $r \to \infty$, Eq. (6.2) yields $g(r) \to 1$ in the same limit as it should. If $f_c = 0$, Eq. (6.2) transforms to (I.6.3). The function $g(r)$ of (6.2) is spherically symmetric, which gives $p_0 = 0$ on account of (3.5) and (3.15). Hence, in the case considered in the present section we have a condensate phase without superfluidity.

We turn now to Eq. (3.17) wherein Eq. (6.2) is substituted. The integrals of $g_I(r)$ will be the same as in I, in particular, (I.6.5) still holds for $g_I(r)$. We are coming to the contribution to the integrals which is due to the first term in (5.1) denoted below as $g_c(r)$. In Appendix D it is shown that in the instance of the potential (I.6.1) one gets

$$\int_0^\infty f(r) K(r) \frac{\partial g_c}{\partial r} dr = \frac{\rho_c \hbar^2 f(a)}{\rho m a^2}. \tag{6.3}$$

If one places $g_c(r)$ instead of $\partial g_c/\partial r$ in an integral of the type (6.3), the method of Appendix D will permit one to show readily that the resulting integral will vanish. Combining (6.3) and (I.6.5) one obtains

$$\int_0^\infty r^3 K(r) \frac{\partial g}{\partial r} dr = \frac{a\hbar^2}{m}\left[f_c + (1 - f_c) H(\bar{\tau})\right], \tag{6.4}$$

where $\bar{\tau} = a^2 m\tau/\hbar^2$ as in (I.6.5). All of these allow Eq. (3.17) to be reduced to the form

$$2\theta \frac{\partial \tilde{\tau}}{\partial \theta} + 3\rho \frac{\partial \tilde{\tau}}{\partial \rho} - 2\tilde{\tau} = \frac{4\pi a \hbar^2 \rho}{3m}\left\{ f_c + (1 - f_c) H(\bar{\tau}) - \frac{a^2 m\theta}{\hbar^2} H'(\bar{\tau}) \frac{\partial \tilde{\tau}}{\partial \theta} \right.$$

$$\left. - \left[1 - H(\bar{\tau}) + \bar{\tau} H'(\bar{\tau})\right]\theta \frac{\partial f_c}{\partial \theta} \right\}, \tag{6.5}$$

where $H(\xi)$ is given by Eq. (I.6.6) with the upper sign, the prime over $H'(\xi)$ denotes differentiation with respect to the argument and, by (3.10),

$$\bar{\tau} = \frac{a^2 m \tilde{\tau}}{\hbar^2 (1 - f_c)}. \tag{6.6}$$

The next step in our study is to elucidate the possibility of bifurcation of a solution with $f_c > 0$ off the high-temperature solution. It is instructive to proceed first on Eq. (6.5) without resorting to the general argumentation of Sec. 3. That equation is rather complex for general analysis. For this reason we restrict ourselves to terms in (6.5) linear in $a$, observing that next terms are of the order $a^5$ in view of (I.6.7)[3]. Then Eq. (6.5) becomes rather simple:

$$2\theta\frac{\partial\tilde{\tau}}{\partial\theta}+3\rho\frac{\partial\tilde{\tau}}{\partial\rho}-2\tilde{\tau} = \frac{4\pi a\hbar^2\rho}{3m}\left(2+\theta\frac{\partial f_c}{\partial\theta}-f_c\right). \tag{6.7}$$

This last equation is identical in form to (B.5), and its general solution for arbitrary $f_c(\theta,\rho)$ is analogous to (B.7). The function $\Phi(\theta/\rho^{3/2})$ is found from the requirement that, at $a = 0$ and $f_c = 0$, the quantity $\tilde{\tau}$ should go over into $\tau$ of an ideal gas, that is, into $\tau$ defined by (I.5.25) and investigated in Appendix G of I (terms with $\theta_0$ which do not contain $f_c$ can be incorporated into $\Phi(\theta/\rho^{3/2})$). If $\tau$ of the ideal gas is denoted as $\tau_0$, there results

$$\tilde{\tau} = \tau_0 + \frac{8\pi a\hbar^2\rho}{3m} + \frac{2\pi a\hbar^2\rho}{3m\sqrt{\theta}}\int_{\theta_0}^{\theta}\tilde{f}_c\left(x,\rho\frac{x^{3/2}}{\theta^{3/2}}\right)\frac{dx}{\sqrt{x}}, \tag{6.8}$$

where $\theta_0$ is arbitrary and

$$\tilde{f}_c(\theta,\rho) = \theta\frac{\partial f_c}{\partial\theta}-f_c = \theta^2\frac{\partial}{\partial\theta}\frac{f_c(\theta,\rho)}{\theta}. \tag{6.9}$$

We need $E$ and $p$ for what follows. These quantities can be calculated by (3.8) and (3.16) with the help of (6.4), so that (cf. (I.6.18))

$$E = \tfrac{3}{2}\tilde{\tau}N, \tag{6.10}$$

$$p = \rho\tilde{\tau} + \frac{2\pi a\hbar^2}{3m}\rho^2\left[f_c+(1-f_c)H(\overline{\tau})\right]. \tag{6.11}$$

The free energy $F$ may be computed with use made of (3.18) or (3.19). It is easier, however, to proceed from (B.1) upon defining $s_0(\rho)$ so that one shall arrive at the free energy of an ideal Bose gas $F_{\text{id}}$ when $a = 0$ and $f_c = 0$. Eqs. (6.10), (6.8) and use of (B.9) give then

$$F = F_{\text{id}} + \frac{4\pi a\hbar^2\rho}{m}N + \frac{\pi a\hbar^2\rho}{m\sqrt{\theta}}N\int_{\theta_0}^{\theta}\left[\frac{1}{\sqrt{x}}f_c\left(x,\rho\frac{x^{3/2}}{\theta^{3/2}}\right)-\frac{\sqrt{x}}{\theta_0}f_c\left(\theta_0,\rho\frac{x^{3/2}}{\theta^{3/2}}\right)\right]dx. \tag{6.12}$$

We note in passing that, at $f_c = 0$, from (6.12) it follows (I.6.27) and (I.6.28). Therefore, Eqs. (I.6.27) and (I.6.28) can be readily obtained from (6.7) with $f_c = 0$ and $\tilde{\tau} = \tau$ without using the rather involved formulae of (I.6.16) and (I.6.17) or the expansions of (I.6.21) and (I.6.22).

Let us assume that the temperature $\theta$ is so close to $\theta_0$ that the variation of $f_c$ in the integral of (6.12) can be neglected within the interval from $\theta_0$ to $\theta$. Then $f_c$ can be brought out from the sign of integration, whereupon the integral is readily evaluated to give

---

[3] As shown in I, such an approximation is equivalent to the one used in the method of pseudopotentials [16].

$$F = F_{id} + \frac{4\pi a\hbar^2 \rho}{m} N + \frac{2\pi a\hbar^2 \rho}{3m\theta_0 \sqrt{\theta}} N\left(\sqrt{\theta_0} - \sqrt{\theta}\right)^2 \left(2\sqrt{\theta_0} + \sqrt{\theta}\right) f_c(\theta_0, \rho). \quad (6.13)$$

From this it follows immediately that any $f_c > 0$ increases the free energy. The same conclusion may be drawn directly from (6.12) on the assumption that the temperature $\theta_0$ is so high that $f_c(\theta_0, \rho) = 0$.

As a result, we see that the free energy is a minimum always at $f_c = 0$, and the bifurcation discussed in Sec. 3 does not occur. The general formulae of Sec. 3 allow one to show that the same remains true not only for small $a$ but also for any $a$. To this end we turn to Eq. (3.20) that, upon substituting (6.4) and (6.6) (in the present case, $\int_0^\infty r^2 K g\, dr = 0$ [1]), takes the form

$$\delta F = \frac{2\pi a\hbar^2 \rho}{m} N \int_1^{\sqrt{\theta_0/\theta}} \left[H(\bar{\tau}) - 1 - \bar{\tau}H'(\bar{\tau})\right] \delta f_c\left(\theta\xi^2, \rho\xi^3\right) d\xi. \quad (6.14)$$

Numerical calculation shows, however, that always $H(\xi) - 1 - \xi H'(\xi) > 0$ for any $\xi$. At small $\xi$, this follows from (I.6.7). For this reason, Eq. (6.14) indicates that the minimum of $F$ occurs only when $f_c = 0$. As to Eq. (3.21) that now assumes the form

$$H(\bar{\tau}) - 1 - \bar{\tau}H'(\bar{\tau}) = 0, \quad (6.15)$$

it can not be satisfied.

The result just obtained shows that no solution branches off from the high-temperature solution in the case studied in the present section. This holds down to the point at which the stability criterion $(\partial p/\partial V)_\theta < 0$ breaks down. Afterwards there emerges a state investigated in Sec. 4. In the case under study, the equations of (4.5) have the form of (I.6.26). When Eq. (I.6.26) was derived from (I.6.18), a relation between $\theta_1$ and $\rho_1$ was first obtained in the following parametric form

$$\sqrt{2m\theta_1} = -\frac{\pi\hbar}{2aG_{-1}(\beta_1)}, \qquad \rho_1 = \omega\theta_1^{3/2} G_0(\beta_1), \quad (6.16)$$

$\beta_1$ being the parameter. Eqs. (I.6.25) and (I.6.26) follow next upon expanding (6.16) in powers of $n = \pi a^3 \rho$ and retaining only first terms. At the same time, corrections to (6.16) are of the order $n^2 \propto a^6$. For this reason, hereafter we shall utilize the equations of (6.16) that are more exact than Eq. (I.6.26). It may be recalled that Eqs. (6.7)-(6.9) and (6.12) are correct to order $a^5$.

According to Eqs. (I.6.27) and (I.6.28), in the high-temperature phase we have, dropping terms in $a^5$,

$$E_h = E_{\text{id}} + \frac{4\pi a\hbar^2 \rho}{m} N = \frac{G_1(\beta)}{G_0(\beta)}\theta N + \frac{4\pi a\hbar^2 \rho}{m} N, \tag{6.17}$$

$$p_h = p_{\text{id}} + \frac{4\pi a\hbar^2 \rho^2}{m} = \frac{2G_1(\beta)}{3G_0(\beta)}\theta\rho + \frac{4\pi a\hbar^2 \rho^2}{m}, \tag{6.18}$$

wherein the parameter $\beta$ is defined by Eq. (I.5.25) which, on putting $\alpha = -\beta^2$, is

$$\rho = \omega \theta^{3/2} G_0(\beta). \tag{6.19}$$

It may be noted in passing that this last equation coincides with the second equation of (6.16).

Inserting all of these into (4.7) yields

$$\varphi_1(\theta) = \omega\theta^{3/2}\left[\frac{2}{3}G_1(\beta_1) + \frac{4\pi a\hbar^2}{m}\omega\theta^{1/2}G_0^2(\beta_1)\right], \tag{6.20}$$

$$\varphi_2(\theta) = -\frac{4\pi a\hbar^2}{m}\omega\theta^{3/2}G_0(\beta_1), \tag{6.21}$$

in which the parameter $\beta_1$ is defined by the first equation of (6.16) with $\theta_1 = \theta$. Now the expressions of (4.4) give $E$ and $p$ at $\theta \leq \theta_1$. At the same time, we have (6.10), so that

$$\tilde{\tau} = \frac{2\omega}{3\rho}\theta^{5/2}G_1(\beta_1) + \frac{8\pi a\hbar^2}{3m}\omega\theta^{3/2}G_0(\beta_1)\left[\frac{2}{\rho}\omega\theta^{3/2}G_0(\beta_1) - 1\right]. \tag{6.22}$$

To the same approximation, Eq. (6.11) assumes the form

$$p = \rho\tilde{\tau} + \frac{4\pi a\hbar^2}{3m}\rho^2\left(1 - \frac{f_c}{2}\right). \tag{6.23}$$

Seeing that $p$ is known by virtue of (4.4), we find from this

$$f_c = 2\left[1 - \frac{\omega}{\rho}\theta^{3/2}G_0(\beta_1)\right]^2. \tag{6.24}$$

On the line (6.16), this quantity vanishes as it should.

According to (6.24), the condensate fraction $f_c$ increases as the temperature $\theta$ is lowered from $\theta_1$ and eventually reaches the value $f_c = 1$ whereas values $f_c > 1$ make no physical sense. Hence, there is another special point $\theta = \theta_2 \neq 0$ at which $f_c = 1$. All quantities pertaining to this last point will be labelled by the subscript 2. The equation of the $f_c = 1$ line on the phase plane follows from Eq. (6.24) with $f_c = 1$ and the first equation of (6.16). The resulting equation of this line is given by

$$\rho_2 = \left(2 + \sqrt{2}\right)\omega\theta_2^{3/2}G_0(\beta_2), \quad \sqrt{2m\theta_2} = -\frac{\pi\hbar}{2aG_{-1}(\beta_2)}, \tag{6.25}$$

where the parameter is denoted as $\beta_2$ in order to differentiate it from the parameters of (6.16) and (6.17)–(6.19). When $a$ is small, $\beta_2$ is small too, which is seen from (I.F.5). In this case, (6.25) is transformed into the following explicit form

$$\theta_2 = \left[ \frac{\rho_2}{(2+\sqrt{2})\zeta_3 \omega} \right]^{2/3}. \tag{6.26}$$

It is instructive to note that, to the same approximation, (6.16) gives $\theta_1 = (\rho_1/\zeta_3 \omega)^{2/3}$.

It should be pointed out that the above derivation of Eqs. (6.25) and (6.26) is not rigorous because Eqs. (6.23) and (6.24) become invalid at $f_c \approx 1$. As $f_c \to 1$, from (6.6) it follows that $\tilde{\tau} \to \infty$ regardless of the value of $a$, and one cannot expand in powers of $a$ as above. When $\tilde{\tau} \to \infty$, one must use the asymptotic formula according to which $H(\xi) \to \xi$ as $\xi \to \infty$ (see I). Then, at $f_c = 1$ Eq. (6.11) yields

$$p = \rho \tilde{\tau} + \frac{2\pi a \hbar^2}{3m} \rho^2 \left( 1 + \frac{a^2 m}{\hbar^2} \tilde{\tau} \right). \tag{6.27}$$

This last expression coincides with (6.23) at $f_c = 1$ provided one discards the term in $a^2$ in the parentheses. This indicates that the equations of (6.25) are correct to order $a^2$ (not inclusive) while Eqs. (6.23) and (6.24) at $f_c \approx 0$ and Eqs. (6.17)–(6.22) too are correct to order $a^5$. To investigate the behaviour of $f_c$ in the neighbourhood of the point $\theta = \theta_2$ it is necessary to know subsequent terms in the asymptotic expansion of $H(\xi)$ as $\xi \to \infty$[4].

When $\theta < \theta_2$, we have $f_c = 1$. To obviate any misunderstanding it should be emphasized that the equality $f_c = 1$, i.e. $\rho_c = \rho$, does not imply that all particles have dropped back to zero level and the thermal motion has ceased. As stressed in I, the quantity $n_s\left(\varepsilon_\nu^{(s)}\right)$ that figures in (2.1) is no occupation number of the level $\varepsilon_\nu^{(s)}$ for there are no single-particle levels in an interacting system, so that one cannot speak of condensation of particles in a level. The particles that form the condensate are in a particular state and at the same time, when $\theta > \theta_2$, they take part in the thermal motion together with other particles as pointed out when discussing (3.6). The thermal motion does not come to an end even below $\theta_2$ when all the particles find themselves in the particular state.

Let us investigate the properties of the system below $\theta_2$. To obtain the form Eq. (6.5) assumes at $f_c = 1$ we pass to the limit as $\tilde{\tau} \to \infty$ according to (6.6). Considering that $H(\xi) \to \xi$ when $\xi \to \infty$, Eq. (6.5) becomes

$$2\theta \frac{\partial \tilde{\tau}}{\partial \theta} + 3\rho \frac{\partial \tilde{\tau}}{\partial \rho} - 2\tilde{\tau} = \frac{4\pi a \hbar^2 \rho}{3m} \left( 1 + \frac{a^2 m}{\hbar^2} \tilde{\tau} - \frac{a^2 m}{\hbar^2} \theta \frac{\partial \tilde{\tau}}{\partial \theta} \right). \tag{6.28}$$

The noteworthy fact is that this last equation is exact as distinct from Eq. (6.7). Upon solving Eq. (6.28) by the method of characteristics one gets

---

[4] To find out analytically the behaviour of $H(\xi)$ when $\xi \to \infty$ is not a simple matter. A numerical calculation permits us to suggest that $H(\xi) \to \xi + \sqrt{\pi \xi} + 2/3 + O(1/\xi)$ as $\xi \to \infty$.

$$\tilde{\tau} = \rho^{2/3} e^{4\pi a^3 \rho/9} \left[ \Phi\left(\frac{\theta}{\rho^{2/3}} e^{-4\pi a^3 \rho/9}\right) + \frac{4\pi\hbar^2}{3m} v\left(a\rho^{1/3}\right) \right], \tag{6.29}$$

where

$$v(x) = \int_0^x \exp\left(-\tfrac{4}{9}\pi\xi^3\right) d\xi. \tag{6.30}$$

It will be noted the following properties of $v(x)$

$$v(x) = x - \frac{\pi}{9}x^4 + \frac{8\pi^2}{567}x^7 + \ldots \quad \text{as } x \to 0, \tag{6.31}$$

$$v(x) = \frac{\Gamma(1/3)}{(12\pi)^{1/3}} - \frac{3}{4\pi x^2}\exp\left(-\tfrac{4}{9}\pi x^3\right) + \ldots \quad \text{as } x \to \infty. \tag{6.32}$$

In order to find the arbitrary function $\Phi(x)$ of (6.29) we equate (6.29) and (6.22) on the line (6.25). Omitting the intervening steps of a tedious but straightforward manipulation we present the expression for $E$ obtained by (6.10) after this, together with the expression for $p$ resulting from (6.27):

$$\frac{E}{N} = \zeta_5 \frac{\omega}{\rho} \theta^{5/2} e^{-2\pi a^3 \rho/3} + \frac{9\zeta_3^2 \theta}{2\sqrt{2}\pi^2}(\theta X - 1)\left[1 - A_1(\theta X - 1) + \ldots\right]$$

$$+ \frac{2\pi\hbar^2}{m}\rho^{2/3} v\left(a\rho^{1/3}\right) e^{4\pi a^3 \rho/9}, \tag{6.33}$$

$$p = \left(1 + \tfrac{2}{3}\pi a^3 \rho\right)\left\{\tfrac{2}{3}\zeta_5 \omega \theta^{5/2} e^{-2\pi a^3 \rho/3} + \frac{3\zeta_3^2 \theta}{\sqrt{2}\pi^2}\rho\theta(\theta X - 1)\left[1 - A_1(\theta X - 1) + \ldots\right]\right.$$

$$\left. + \frac{4\pi\hbar^2}{3m}\rho^{5/3} v\left(a\rho^{1/3}\right) e^{4\pi a^3 \rho/9}\right\} + \frac{2\pi a\hbar^2 \rho^2}{3m} \tag{6.34}$$

with

$$X = \left[\frac{(2+\sqrt{2})\zeta_3 \omega}{\rho}\right]^{2/3} e^{-4\pi a^3 \rho/9}, \qquad A_1 = \frac{11 + 7\sqrt{2}}{4 + 2\sqrt{2}} + \frac{9\zeta_1\zeta_3}{4\pi^2}. \tag{6.35}$$

The expression in the square brackets in (6.33) and (6.34) represents first terms of an expansion in powers of $\theta X - 1$ which can be obtained completely if all terms of the expansions (I.6.21) and (I.6.22) for the high-temperature phase are known. It should be added that $\theta X = 1$ on the line (6.25) upon discarding terms of order $a^2$ (this is seen also directly from (6.26)).

For the purpose of constructing resultant isotherms, we calculate the derivative $(\partial p/\partial \rho)_\theta$ on the line (6.25). Since $\theta X \approx 1$ on that line, Eq. (6.34) entails, in a first approximation in $a$,

$$\left(\frac{\partial p}{\partial \rho}\right)_\theta = -\frac{\sqrt{2}\zeta_3^2\,\theta_2}{\pi^2}\left\{1 - \frac{2\sqrt{2}}{3}\left[\frac{\pi(1+\sqrt{2})}{\zeta_3^2}\right]^{2/3}\pi a\rho_2^{1/3}\right\} \approx -\frac{\sqrt{2}\,\zeta_3^2\,\theta_2}{\pi^2}\left(1 - 2.55\,n_2^{1/3}\right). \quad (6.36)$$

The derivative is negative if $n_2 = \pi a^3 \rho_2$ is small. A schematic isotherm that summarizes the results of the present section for small $n$ is sketched in figure 2 by the solid curve where use is made of the specific volume $v = 1/\rho$ instead of the density $\rho$. Eq. (6.34) (see the last term in the braces) on account of (6.32) shows that the pressure rises exponentially as $\rho \to \infty$, i.e., as $v \to 0$.

The isotherm plotted in figure 2 (the solid line) is typical of a first-order phase transition. The transition point may be found by making use of the Maxwell construction [16], for example. We cannot, however, do this in the present paper. The reason is that Eq. (6.34) holds only not far from the point $v = v_2$. In order to compute the pressure on the ascending part of the curve at $v < v_2$, it is necessary to know the complete series in the square brackets of (6.34). When $\theta X - 1$ is arbitrary, the series is not an expansion in powers of a small parameter (the smallness of $a$ does not play an essential role for the series) and can be summed up only if use is made of the exact solution for the high-temperature phase according to the remark following (6.35).

If $n_2 > 6.04 \times 10^{-2}$, then $(\partial p/\partial v)_\theta < 0$ at $v = v_2$ according to (6.36), and the isotherm at $v \leq v_2$ has the form of the broken curve in figure 2. The isotherm can be continued to volumes $v > v_2$ having the derivative $(\partial p/\partial v)_\theta < 0$ at least when $v \approx v_2$. The situation will correspond to a first-order phase transition as before. The case $n_2 = 6.04 \times 10^{-2}$ where $(\partial p/\partial v)_\theta = 0$ at $v = v_2$ is worthy of special comment. Although in this case the isotherm is similar to isotherms for a first-order transition, the actual situation is different. When $v_1 < v < v_2$, there exists a single-phase system, but not a two-phase system. In fact, we have two continuous phase transitions, at $v = v_1$ and $v = v_2$. It is not excluded that, if a more exact approximation is used, the derivative $(\partial p/\partial v)_\theta$ will become zero at $v = v_2$ even when $n_2 > 6.04 \times 10^{-2}$; and the two continuous phase transitions will occur as well. It is not clear, however, whether the situation with such two transitions can be observed in reality. Although the value $n_2 = 6.04 \times 10^{-2}$ is not large, the triplet correlations that have been discarded may play an important part in the condensed region even if $n$ is rather small (the importance of correlations is discussed below). The triplet correlations may substantially affect the isotherm in the vicinity of the point $v = v_2$ and even the horizontal part of the isotherm altogether. As to the isotherm represented by the solid line in figure 2, the exact behaviour of the isotherm between $v = v_1$ and $v = v_2$ does not play an essential role because we have a first-order transition in this case.

The Bose condensation in a hard-sphere system was studied earlier by the method of pseudopotentials [23,24] (see also Refs. [12,16]). In those studies, in a first approximation in *a* the condensation appears to be a second-order transition, and therefore those isotherms differ essentially from the ones plotted in figure 2. At the same time, in Ref. [24] it is noted that it is not possible to state exactly the order of the transition for a finite *a*, whereas in Ref. [23] (see also [12]) the authors argue that, even in the first approximation in *a*, the model considered there leads to a first-order transition if use is made of the grand canonical ensemble. Once the influence of higher-order terms is so essential that owing to them even the order of the transition changes, account taken of them may modify appreciably other results, too, obtained in Refs. [23,24] by considering only terms of first order in *a*. It may be recalled that we took into account several orders in *a* from the very beginning (see the remark following (6.16)).

In concluding the section let us make some remarks upon the results we have obtained here. From (6.26) it follows that the point $\theta = \theta_2$ (or $v = v_2$) should exist at $a = 0$ as well, that is to say, it would seem that such a point should occur even in an ideal gas. Besides, Eqs. (6.33) and (6.34) do not, upon putting $a = 0$, go over into the corresponding equations for an ideal gas below the condensation point. In fact, the situation is not so simple and is due to the following. Eq. (6.2) shows that the condensate fraction induces long-range spatial correlations decaying as $1/r$ while the last term in (6.2) relevant to the normal fraction is short-ranged (see the discussion following (I.6.3)). At the same time, both the fractions are connected by the common potential $U_2(r)$ that in turn depends upon them. For this reason the long-range correlations in the condensate fraction must cause analogous correlations in the normal fraction. From this it follows that the present approximation consisting in the neglect of triplet correlations in calculation of $U_2(r)$ is not good for the condensate phase even if $a \to 0$. The influence of the long-range correlations will lead to corrections in Eq. (6.23) for the pressure even in a first approximation in *a*. When obtaining (6.24), the factor *a* cancels out. As a result, we shall arrive at another expression for $f_c$ even in the limit $a \to 0$ and consequently at another expression for $\theta_2$ instead of (6.26). From (6.6) it follows that $\bar{\tau} \to \infty$ as $f_c \to 1$; therefore, all details of the structure are of importance even if $a \to 0$, which is due to the fact that all particles pass into the exceptional state. This results in the terms of (6.33) and (6.34) weakly dependent on *a* whose form will, of course, change if the triplet correlations are taken into account. The foregoing shows that properties of the condensate phase are different even in the limit as the interaction potential tends to zero, that is to say, there is no unique limit for the properties[5]. Probably, this fact explains the difficulties the authors of Refs. [23,24] had met with when determining the order of the transition. As to our approach, a more consistent account of the correlations may change only quantitative details while not affecting the qualitative aspects of the problem.

We see that the physical reason that leads to the foregoing peculiarities of the condensate phase is the presence of long-range correlations which are due to the fact that the condensate particles are in an exceptional state. This state and thereby the value of $f_c$ are influenced substantially by details of the interaction and, quite possibly, by peculiarities of the structure of the system. It is logical to suppose that such an influence on $f_c$ remains also in the case where $\mathbf{p}_0 \neq 0$. Eq. (3.15) shows that then $p_0$, too, will be considerably affected by the characteristics of the structure. In the concluding section we shall see that $p_0$ determines the critical velocity. Consequently, the critical velocity should depend noticeably upon conditions of the experiment, impurities in the liquid, inhomogeneities, etc., which may account for discrepancies between the values of critical velocities in helium observed in different experiments [25].

## 7. Weakly interacting systems

The second example considered in I is a weakly interacting system in which the interaction potential admits a Fourier transform. The results of Sec. 7 of I for this example give terms of orders $\lambda^0$ and $\lambda$ in thermodynamic quantities, the procedure being factually analogous with the method considered in Appendix C. According to Appendix C, the condensate phase in this case will not be superfluid ($\mathbf{p}_0 = 0$) as in Sec. 6.

In the approximation in question, if a formula contains $g(r)$ multiplied by $K(r)$, one should replace $g$ by its value at $K \equiv 0$. If $K \equiv 0$, then $U_2 \equiv 0$ too, in which case Eq. (2.25) yields $u_2 = 1$. The second term in (5.1) at $U_2 \equiv 0$ may be found with the help of (I.7.1) (the upper sign) upon taking account of the fact that the normalization factor $A$ is now given by (3.1) (cf. (6.2)); hence, the required form of $g$ is

$$g(r,\theta,\rho) = 1 + (1 - f_c) \exp\left[-\frac{mr^2}{\hbar^2}\tau_0(\theta,\rho)\right] \qquad (7.1)$$

with $\tau_0(\theta,\rho)$ defined by (I.5.25). Eq. (3.17) is not needed in the present approximation as long as we have the explicit expression for $g$ in terms of $f_c$, $\theta$ and $\rho$.

---

[5] As mentioned in Sec. 8 of I, the case of an ideal gase ($a = 0$) is not so simple. Recall also that we could not find $f_c$ when deriving (4.9).

We now investigate the question as to whether a solution with $f_c > 0$ can bifurcate off the high-temperature solution considered in Sec. 7 of I. We introduce (7.1) into (3.19) and choose $\theta, \rho, \theta_0$ and $\rho_0$ such that the second integral in (3.19) shall vanish. All terms that do not contain $f_c$ will be denoted by $F_h$ upon assuming that $f_c$ does not enter in $F(\theta_0, \rho_0)$ (this implies the temperature $\theta_0$ to be sufficiently high according to the discussion following (3.19)). We make use of the following property of $\tau_0(\theta, \rho)$ which is a consequence of (I.5.22):

$$\tau_0(\theta\xi^2, \rho\xi^3) = \xi^2 \tau_0(\theta, \rho). \tag{7.2}$$

Upon integrating the term with $dK/dr$ in (3.19) by parts we obtain finally

$$F = F_h + \frac{\rho}{2} N \int_1^{\sqrt{\theta_0/\theta}} \left[ K_1\left(\xi^2 \tau_0\right) - \frac{2m\tau_0}{\hbar^2} \xi^2 K_2\left(\xi^2 \tau_0\right) \right] f_c\left(\theta\xi^2, \rho\xi^3\right) d\xi, \tag{7.3}$$

where $\tau_0 = \tau_0(\theta, \rho)$ and the notation of (I.7.7) is employed.

Let us analyse (7.3) in the limit of high temperatures when $\tau_0 \to 0 \to \infty$ (of course, $\theta < \theta_0$). If $\xi \to \infty$, the integrals of (I.7.7) behave asymptotically as follows

$$K_1(\xi) \to \left(\frac{\pi\hbar^2}{m\xi}\right)^{3/2} K(0), \qquad K_2(\xi) \to \frac{3\hbar^2}{2m\xi} \left(\frac{\pi\hbar^2}{m\xi}\right)^{3/2} K(0). \tag{7.4}$$

These allow one to reduce (7.3) in this limit to

$$F = F_h - \rho K(0) N \left(\frac{\pi\hbar^2}{m\theta}\right)^{3/2} \int_1^{\sqrt{\theta_0/\theta}} f_c\left(\theta\xi^2, \rho\xi^3\right) \frac{d\xi}{\xi^3}. \tag{7.5}$$

If the core of the potential is attractive ($K(0) < 0$), the value $f_c = 0$ is thermodynamically preferable according to (7.5). However, if the core is repulsive ($K(0) > 0$), it is the value $f_c > 0$ that is preferable. This last result contradicts the common sense because we imply temperatures that may be arbitrarily high even classical; the result disagrees also with the results of the preceding section. This points out that the week-coupling approximation is unsuitable for problems of the type under study; in these problems the impenetrable core plays an essential role. The physical reason for this is quite understandable. If we put $f_c = 0$ in (7.1), in the limit as $\tau_0 \to \infty$ we obtain $g \to 1$ for all $r$ except for $r = 0$. At $r = 0$, always $g = 2$ regardless of $\tau_0$. Consequently, even noninteracting particles would have a tendency to coalesce at any temperature, which is devoid of sense at classical temperatures. The tendency may look natural in the case of an attractive core of the potential but the tendency is nonphysical if the core is repulsive. Because of this the week-coupling approximation can make some sense only in the case of the attractive core. To avoid any manifestation of the nonphysical effect of coalescence a hard core should be introduced.

Consideration of a hard core lies beyond the framework of the week-coupling approximation dealt with in the present section (if $K(r) = \infty$ when $r < a$, the integrals $K_1$ and $K_2$ of (I.7.7) do not exist). Although the approximation leads to the nonphysical effects, we can utilize (7.3) as an example in order to explore the possibility of bifurcation. Let $\theta$ and thereby $\tau_0$ be small, in which case $\theta_0/\theta \gg 1$. Then the leading contribution to the integral in (7.3) is made by small values of $\xi$, say $\xi < \xi_0$, for at large $\xi$ both the terms in the square brackets in (7.3) practically cancel each other by virtue of (7.4). At small $\xi$, the first term should dominate owing to the small factor $\tau_0$ in the second term. If this is so, then

$$F \approx F_h + \tfrac{1}{2}\rho K_0 N \int_1^{\xi_0} f_c\left(\theta\xi^2, \rho\xi^3\right) d\xi, \qquad (7.6)$$

upon recalling that $K_1(0) = K_0$ (see I). We see that the second term in (7.6) has reversed the sign with respect to the second term in (7.5). For this reason (see Sec. 3) a continuous transition to the condensate phase would be possible in the present case.

Although the foregoing consideration can be of only academic interest, it does show that the phase transition to the condensate phase may be not only first order as in Sec. 6 but also continuous. To investigate details of such a continuous transition on the basis of (7.1) and (7.3) does not make sense because we may obtain unreasonable results. The week-coupling approximation with account taken of higher orders according to Appendix C might be used in the problem under consideration on condition one could rearrange terms in the relevant series by making use of ideas expounded, for example, in Sec. 6.1 of Ref. [26].

## 8. Discussion and concluding remarks

The present paper shows that within the scope of the approach proposed in I lies also such a phenomenon as superfluidity. However, in contradistinction to the traditional view on superfluidity the presented approach indicates that the phenomenon embodies symmetry breaking in a fluid which is due to formation of a spontaneous stationary flow, the state with the flow being thermodynamically equilibrium while the magnitude of the flow being determined by thermodynamic relations. The new view places superfluidity amongst such well-understandable phenomena as the appearance of a spontaneous magnetization or polarization below the Curie point in a ferromagnet or ferroelectric, the values of the magnetization or polarization being given by thermodynamics (hereafter for brevity we shall refer to ferromagnets only). Insofar as the state with the flow corresponds to thermal equilibrium, no dissipation of energy can occur.

It is important to emphasize that the quantity $\mathbf{p}_0$ which determines the flow appears in the theory quite naturally since there is no reason to put $\varepsilon_{(1)} = 0$ in Eq. (2.13). Thus, no additional argument is needed to explain the existence of the frictionless flow (cf. Introduction). Factually, the subsequent treatment is merely demonstration of the fact that the existence of the flow does nowise contradict equilibrium thermodynamics (in parallel we investigated various aspects of the problem).

Let us turn now to the question of how to elucidate superfluidity physically. We note first that incessant motion without any dissipation of energy is characteristic of quantum mechanics. As an example, we may mention the motion of electrons in an atom, which may manifest itself externally by the existence of an orbital angular momentum $\mathbf{L}$. Recall that not all electrons contribute to the total angular momentum although all the electrons constitute a unified dynamical system. Recall also that an atom, let alone a molecule, can contain up to a hundred electrons[6] and its existence is a purely quantum effect because it could not exist according to classical physics. A superfluid may be regarded as a gigantic atom whose radius is infinite, and in which for this reason the orbital motion is converted to a rectilinear one that is characterized by a linear momentum $\mathbf{P}$ instead of the angular momentum $\mathbf{L}$. The gigantic atom differs from an ordinary one in that the role of electrons is played by neutral atoms which move in a self-consistent field, not all of them contributing to $\mathbf{P}$ (cf. the discussion of Eq. (2.29)). It is clear also why the contribution to the superflow is made only by condensate atoms. According to Sec. 2 these last atoms are in an exceptional state, and given this they are especially affected by specific quantum mechanical laws. Here again we may draw a parallel between ferromagnetism and superfluidity. Ferromagnetism may be regarded as a macroscopic manifestation of the spin angular momentum while superfluidity is a peculiar macroscopic manifestation of the orbital angular momentum. In an ordinary atom both states with $\mathbf{L} \neq 0$ and $\mathbf{L} = 0$ are possible; analogously, the condensate phase may be either superfluid ($\mathbf{p}_0 \neq 0$) or nonsuperfluid ($\mathbf{p}_0 = 0$).

A similar view may be expressed as to explanation of superconductivity. In a superconductor charged particles, electrons, are entrained in the collective quantum motion under discussion, which gives rise to a supercurrent. From this point of view, high-temperature superconductivity is explained probably by a singular structure of relevant compounds, which favours the motion at high temperatures. Inasmuch as the electrons have a spin, their characteristic attribute, in order to consider superconductivity in this way it is necessary first to extend the approach developed in I to systems containing nonzero spin particles, which will be the subject of a subsequent paper.

---

[6] From the viewpoint of computer simulation, a system of 100 particles is a macroscopic system.

In this context the following should be noted. Theories of superconductivity (superfluidity of electrons) based upon an idea that the phenomenon is due to spontaneous currents existing in a condition of thermodynamic equilibrium, which is analogous with the results of the present paper, were proposed as far back as the 1930s (for a review see Ref. [27], a critical analysis of such theories can be found in Ref. [28] as well). A characteristic paper along these lines is that of Landau [29]. Landau assumed that a superconductor contains local saturation currents flowing in different directions and yielding no resultant current in the absence of an external field. His study was not based upon quantum mechanics, being phenomenological in essence and without convincing argumentation. All those theories were not sufficiently founded and were abandoned later. One of the main objections against the theories was the following [28]. Superfluidity was explained by Landau's mechanism [2,3] radically different from the mechanism of spontaneous currents, whereas the mechanism should be universal because superconductivity can be regarded as superfluidity of an electron gas. The present paper suggests an universal mechanism for superfluidity and superconductivity. Besides, the paper provides a firm foundation for the mechanism of spontaneous flows or currents.

The present approach shows that there is a preferred direction specified by the vector $\mathbf{p}_0$, which implies an anisotropy in the velocity distribution. In recent experiments on Bose condensation in a gas of alkali atoms [30,31] one observed such an anisotropy. At the same time a direct comparison of our results with the experimental ones is impossible because in the present paper an idealized case is considered, in which the system occupies an infinite volume, whereas real volumes are always finite.

Let us discuss manifestation of superfluidity in finite systems. In a confined volume flows must close upon themselves. It is quite possible that the volume will break up into cells in each of which the fluid will circulate along closed streamlines, forming a pattern similar to that observed, for example, in Rayleigh-Bénard convection (for a recent review see, e.g., Ref. [32]). Here again an analogy with a ferromagnet is in place: the cells in the superfluid are analogous to domains in the ferromagnet. At the same time there is a great difference. Formation of magnetic domains is energetically profitable because owing to them the magnetic energy of the specimen decreases. Curvature of the flow and formation of the cells in a superfluid will most likely be energetically unfavourable in all respects. Hence in some cases, especially just below the λ point, the cells may not develop and one will observe a metastable condensate phase without superfluidity; in other cases, under appropriate circumstances, a cell pattern may arise.

As to the motion of the superfluid near walls of the vessel we note that atoms of the walls may adjust themselves to the quantum motion since their temperature is low as well. When trying to visualize the flow pattern one may meet with some peculiarities for the superflow does not exert an extra pressure along the streamlines (see the discussion of Eq. (3.14)) as distinct from a flow in an ordinary liquid. The existence of the intrinsic flow explains readily the known fact that helium II crawls up the walls, out of an open vessel (see, e.g., Ref. [16]). This sheds further light on the fountain effect as well.

The present ideas offer also a simple explanation for the existence of a critical velocity. If a superfluid in which the foregoing pattern has formed is subjected to an external action, be it a gradient of pressure or temperature, some of the closed flows will open and the superfluid will partially flow along the direction of the gradient. Here again we have an analogy with a ferromagnet in which an external magnetic field reorients the magnetic moments of the domains. To the saturation magnetization, when the specimen becomes one domain, will correspond a state of the superfluid in which the whole flow will be directed along the gradient. A further increase in the total flow is impossible without breaking the condition of thermodynamic equilibrium, which will result in viscosity. Hence, the critical velocity is determined by the formula $v_c = p_0/m$ according to the discussion following Eq. (2.29), where $p_0$ as a function of the temperature and density is given by Eq. (3.15). In narrow and long capillaries where the creation of closed flows is hindered, superfluidity will manifest itself easier, which conforms to experiment. The reasoning remains practically the same if the formation of the cells is energetically unfavourable: a temperature or pressure gradient will again facilitate creation of a directional flow, and the critical velocity will again be given by $v_c = p_0/m$. In Landau's explanation of superfluidity [2,3] the critical velocity is of the same order of magnitude as the velocity of sound, that is, it is too large (see Introduction); in the present theory the critical velocity is not at all related with the velocity of sound as seen from Eq. (3.15).

If superfluidity is regarded as spontaneous symmetry breaking, the critical velocity may be considered the order parameter vanishing at the transition point. It should be observed that, strictly speaking, in terms of the Landau theory of phase transitions the critical velocity is a secondary order parameter while the primary order parameter is $f_c$ (more precisely $\sqrt{f_c}$).

A word should be said about the vortices. When the velocity of the flow exceeds the critical velocity, the equilibrium superfluid state breaks down, which gives rise to vortices. From this point of view the breakdown of superfluidity is the cause for the creation of vortices. In Feynman's argument [6] this causal chain is reversed; the creation of vortices is the cause and the breakdown of superfluidity is the effect. It should be stressed that the cells discussed above have nothing to do with the vortices. On the other hand, the cells may somehow favour the formation of the vortices when superfluidity breaks down.

Let us discuss now the phases that were called the condensate phases without superfluidity. We revert again to the analogy with an atom. If in the ground state of the atom the orbital angular momentum **L** is zero, an excited state can be such that $\mathbf{L} \neq 0$. In a superfluid the role of the exciting agent can be played by a temperature or pressure gradient. Consequently, the $\mathbf{p}_0 = 0$ state can be converted to a metastable $\mathbf{p}_0 \neq 0$ state. Once a quantum system goes to another state, its subsequent behaviour does not depend upon the exciting agent (if there is no other transition). Therefore, in the emerging state of the superfluid the quantity $\mathbf{p}_0$ will be determined by internal properties of the superfluid, that is to say, by Eq. (3.15). On the whole, the properties of the metastable $\mathbf{p}_0 \neq 0$ state should be analogous with those of a genuine superfluid. The question of lifetime of the metastable state lies beyond the scope of equilibrium statistical mechanics. In the case of a superconductor the role of the exciting agent in question may be played by not only an external voltage but also a contact potential.

## Appendix A. Stress tensor

In this appendix the particles will be labelled by the indices $p,q$ ($1 \leq p,q \leq N$) and the components of the vector $\mathbf{r}_p$ by $i,j,k$ ($1 \leq i,j,k \leq 3$, with summation over repeated indices). Let us denote the second term of (I.2.6) by $\tilde{H}$, put $V^{(e)} = 0$ and write down the time-dependent Schrödinger equation

$$-\frac{\hbar^2}{2m}\sum_{p=1}^{N}\frac{\partial^2 \Psi}{\partial x_{pk}^2} + \left(\tilde{H} - i\hbar\frac{\partial}{\partial t}\right)\Psi(\mathbf{x}_N, t) = 0. \qquad (A.1)$$

At the beginning we follow partly [21, page 68]. We differentiate Eq. (A.1) with respect to $x_{qi}$, multiply it by $x_{qj}\Psi^*(\mathbf{x}_N, t)$ and make use of the complex conjugate of (A.1). This yields

$$-\frac{\hbar^2}{2m}\sum_{p=1}^{N} x_{qj}\left(\Psi^*\frac{\partial^3 \Psi}{\partial x_{qi}\partial x_{pk}^2} - \frac{\partial \Psi}{\partial x_{qi}}\frac{\partial^2 \Psi^*}{\partial x_{pk}^2}\right) + x_{qj}\frac{\partial \tilde{H}}{\partial x_{qi}}|\Psi|^2 - i\hbar x_{qj}\frac{\partial}{\partial t}\left(\Psi^*\frac{\partial \Psi}{\partial x_{qi}}\right) = 0. \quad (A.2)$$

Each term in the sum over $p$ may be represented as

$$\frac{\partial}{\partial x_{pk}}\left[\Psi^{*2}\frac{\partial}{\partial x_{pk}}\left(\frac{x_{qj}}{\Psi^*}\frac{\partial \Psi}{\partial x_{qi}}\right)\right] - 2\delta_{pq}\Psi^*\frac{\partial^2 \Psi}{\partial x_{pi}\partial x_{qj}}.$$

Upon substituting this into (A.2) and summing over $q$ from 1 to $N$, we integrate the resulting expression with respect to all $\mathbf{r}_p$ over the volume of the system. The integral of $\partial[...]/\partial x_{pk}$ can be transformed into a surface integral and gives a contribution due to the boundary of the system. The contribution depends upon the environmental conditions the system is placed in and can be expressed in terms of the virial of external forces applied to the surface of the system. Denoting this last contribution by $\Xi_{ij}$ we have

$$\frac{\hbar^2}{m}\sum_{q=1}^{N}\int \Psi^*\frac{\partial^2\Psi}{\partial x_{qi}\partial x_{qj}}d\mathbf{x}_N + \sum_{q=1}^{N}\int x_{qj}\frac{\partial \tilde{H}}{\partial x_{qi}}|\Psi|^2 d\mathbf{x}_N + \Xi_{ij} - i\hbar\frac{\partial}{\partial t}\sum_{q=1}^{N}\int x_{qj}\Psi^*\frac{\partial \Psi}{\partial x_{qi}}d\mathbf{x}_N = 0. \quad (A.3)$$

We can get the expression for $\Xi_{ij}$ by examining the structure of the second term of (A.3) with account taken of the fact that $-\partial \tilde{H}/\partial x_{pi}$ is the $x_i$-component of the force acting upon a particle at the point $\mathbf{r}_p$. Let $\mathbf{P}$ be the external force exerted on a unit element of the surface of the system. Then

$$\Xi_{ij} = -\int_S x_j P_i \, dS, \quad (A.4)$$

where the integration is extended over the surface of the system. The last integral can be expressed in terms of a space average of the stress tensor $\sigma_{ij}$ [33], so that $\Xi_{ij} = -V\langle\sigma_{ji}\rangle$. For a macroscopically homogeneous medium, we have $\langle\sigma_{ji}\rangle = \sigma_{ji} = \sigma_{ij}$.

The other terms of (A.3) are transformed by using the definition of the density matrices (I.2.1) (cf. (I.3.6)) while the last term will depend on $\partial R_1/\partial t$. As we consider only the stationary part of $R_1$ (see I), this last term is to be discarded. As a result, (A.3) and (A.4) yield

$$\sigma_{ij} = \frac{\hbar^2}{mV}\int\left[\frac{\partial^2}{\partial x_i \partial x_j}R_1(\mathbf{r},\mathbf{r}')\right]_{\mathbf{r}'=\mathbf{r}} d\mathbf{r} + \sigma_{ij}^{(2)} \quad (A.5)$$

with

$$\sigma_{ij}^{(2)} = \int x_{1j}\frac{\partial K(|\mathbf{r}_1-\mathbf{r}_2|)}{\partial x_{1i}}\rho_2(\mathbf{r}_1,\mathbf{r}_2)d\mathbf{r}_1\, d\mathbf{r}_2. \quad (A.6)$$

The integral in (A.6) is inconvenient for evaluation because regions near the surface of the system where $|x_{1j}|$ is large make a contribution proportional to the volume of the system. Therefore, to evaluate the integral a knowledge of $\rho_2(\mathbf{r}_1,\mathbf{r}_2)$ in those regions is required whereas we consider only bulk properties. We can transform (A.6). Taking account of the identity $\partial K(|\mathbf{r}_1-\mathbf{r}_2|)/\partial x_{1i} = -\partial K(|\mathbf{r}_1-\mathbf{r}_2|)/\partial x_{2i}$ and interchanging the indices 1 and 2 gives

$$\sigma_{ij}^{(2)} = -\int x_{2j}\frac{\partial K(|\mathbf{r}_1-\mathbf{r}_2|)}{\partial x_{1i}}\rho_2(\mathbf{r}_2,\mathbf{r}_1)d\mathbf{r}_1\, d\mathbf{r}_2. \quad (A.7)$$

Upon adding (A.6) and (A.7) and using the symmetry of $K(|\mathbf{r}_1-\mathbf{r}_2|)$ and $\rho_2(\mathbf{r}_1,\mathbf{r}_2)$ we obtain

$$\sigma_{ij}^{(2)} = \frac{1}{2}\int(x_{1j}-x_{2j})\frac{\partial K(|\mathbf{r}_1-\mathbf{r}_2|)}{\partial x_{1i}}\rho_2(\mathbf{r}_2,\mathbf{r}_1)d\mathbf{r}_1\, d\mathbf{r}_2. \quad (A.8)$$

In the last integral, $|x_{1j} - x_{2j}|$ does not exceed the range of the interparticle interaction owing to $K(|\mathbf{r}_1-\mathbf{r}_2|)$. Therefore, the regions close to the surface of the system make a contribution that can be neglected if $V \to \infty$.

We perform the replacements $\mathbf{r}_1 - \mathbf{r}_2 = \mathbf{r}$ and $\mathbf{r}_2 = \mathbf{r}'$ and introduce a space average according to

$$\langle \rho_2(\mathbf{r}',\mathbf{r}'+\mathbf{r}) \rangle = \frac{1}{V} \int \rho_2(\mathbf{r}',\mathbf{r}'+\mathbf{r}) d\mathbf{r}'. \qquad (A.9)$$

On placing (A.8) with (A.9) in (A.5) we arrive at

$$\sigma_{ij} = \frac{\hbar^2}{mV} \int \left[ \frac{\partial^2}{\partial x_i \partial x_j} R_1(\mathbf{r},\mathbf{r}') \right]_{\mathbf{r}'=\mathbf{r}} d\mathbf{r} + \frac{1}{2} \int x_j \frac{\partial K(|\mathbf{r}|)}{\partial x_i} \langle \rho_2(\mathbf{r}',\mathbf{r}'+\mathbf{r}) \rangle d\mathbf{r}. \qquad (A.10)$$

This formula holds for gases and liquids as well as for crystals. For spatially homogeneous media when $\rho_2 = \rho_2(\mathbf{r}_1 - \mathbf{r}_2)$, we obtain (3.7).

### Appendix B. Formulae for the free energy

The free energy can be calculated similarly to (I.5.28). Let us obtain a general formula. Integrating (I.5.27) gives

$$F(\theta,\rho) = -\theta \int_{\theta_0}^{\theta} E(x,\rho) \frac{dx}{x^2} - \theta N s_0(\rho) \qquad (B.1)$$

with an arbitrary constant $\theta_0$. In order to find the function $s_0(\rho)$ we differentiate (B.1) with respect to $\rho$ and take account of (I.5.29) and also of (3.13) written as

$$\frac{\rho^2}{N} \left( \frac{\partial E}{\partial \rho} \right)_\theta = -\theta^2 \left( \frac{\partial}{\partial \theta} \frac{p}{\theta} \right)_\rho. \qquad (B.2)$$

Then the integral containing $\partial E/\partial \rho$ is readily evaluated and we get

$$\frac{ds_0}{d\rho} = -\frac{1}{\theta_0 \rho^2} p(\theta_0,\rho), \qquad s_0(\rho) = -\frac{1}{\theta_0} \int_{\rho_0}^{\rho} p(\theta_0,y) \frac{dy}{y^2} + \bar{s}_0, \qquad (B.3)$$

where $\rho_0$ and $\bar{s}_0$ are arbitrary constants. The first equation of (B.3) corresponds to (I.5.30) with reference to (I.5.17). The constant $\bar{s}_0$ is expressed via $F(\theta,\rho)$ at $\theta = \theta_0$ and $\rho = \rho_0$, which gives (3.18). By integrating (I.5.29), likewise one can obtain another formula,

$$F(\theta,\rho) = N \int_{\rho_0}^{\rho} p(\theta,y) \frac{dy}{y^2} - \theta \int_{\theta_0}^{\theta} E(x,\rho_0) \frac{dx}{x^2} + \frac{\theta}{\theta_0} F(\theta_0,\rho_0). \qquad (B.4)$$

It may be noted that this last formula can be also derived from (3.18) if one interchanges $\rho$ and $\rho_0$ as well as $\theta$ and $\theta_0$.

Let us obtain still another formula for the free energy. To this end we find first an equation for $E(\theta,\rho)$ by solving (3.16) for $\tilde{\tau}$ and placing this in (3.17), so that

$$2\theta \frac{\partial E}{\partial \theta} + 3\rho \frac{\partial E}{\partial \rho} - 2E = 2\pi N\, M(\theta,\rho) \tag{B.5}$$

with

$$M(\theta,\rho) = \rho\theta^2 \int_0^\infty dr\, r^2 \left[2K(r) + r\frac{dK}{dr}\right] \frac{\partial}{\partial \theta} \frac{\bar{g}(r,\theta,\rho)}{\theta}. \tag{B.6}$$

We solve (B.5) by the method of characteristics (cf. (I.6.9) and (I.7.3)):

$$E(\theta,\rho) = \rho^{2/3}\Phi\!\left(\theta/\rho^{2/3}\right) + \pi\theta N \int_{\theta_0}^{\theta} M\!\left(x, \rho \frac{x^{3/2}}{\theta^{3/2}}\right) \frac{dx}{x^2}, \tag{B.7}$$

where $\Phi(x)$ is a function to be determined. The equivalence of (B.7) and (B.5) can be easily verified by substituting (B.7) into (B.5). The function $\Phi(x)$ can be found if one knows the energy $E(\theta,\rho)$ at $\theta = \theta_0$ and arbitrary $\rho$:

$$\Phi(x) = \frac{x}{\theta_0} E\!\left(\theta_0, \frac{\theta_0^{3/2}}{x^{3/2}}\right). \tag{B.8}$$

We insert (B.7) with (B.8) into (3.18). It is convenient to choose $\theta_0$ in (B.7) and (3.18) to be the same. The integral of the second term of (B.7) is transformed with the help of the following identity that can be proven easily enough for arbitrary functions $f_1(x)$ and $f_2(x,y)$

$$\int_{\theta_0}^{\theta} dy\, f_1(y) \int_{\theta_0}^{y} dx\, f_2(x, \rho \frac{x^{3/2}}{y^{3/2}}) = \frac{2}{3}\rho^{2/3} \int_{\rho(\theta_0/\theta)^{3/2}}^{\rho} \frac{dy}{y^{5/3}} \int_{\theta_0}^{\theta(y/\rho)^{2/3}} dx\, x\, f_1\!\left(x\frac{\rho^{2/3}}{y^{2/3}}\right) f_2(x,y). \tag{B.9}$$

If, upon putting $f_1(x) = 1/x$ and $f_2(x,y) = M(x,y)/x^2$, we substitute (B.6) here, the integral over $x$ is calculated at once. We simplify the arguments of the functions by changing the variables of integration and transform the term that contains $\bar{g}(r,\theta_0,y)$ with the aid of the relation

$$\frac{2}{N} E(\theta,\rho) - \frac{3}{\rho} p(\theta,\rho) = 2\pi\rho \int_0^\infty dr\, r^2 \left[2K(r) + r\frac{dK}{dr}\right] \bar{g}(r,\theta,\rho), \tag{B.10}$$

which follows from (3.8) and (3.16). As a result, we obtain (3.19).

It may be noted that, if by analogy with (B.5) one derives the equation for $p(\theta,\rho)$ and makes use of (B.4), one obtains, instead of (3.19),

$$F(\theta,\rho) = 2\pi\rho N \int_0^\infty dr\, r^2 \left(2K + r\frac{dK}{dr}\right) \int_1^{(\rho_0/\rho)^{1/3}} d\xi\, \bar{g}(r,\theta\xi^2,\rho\xi^3) + \theta \int_{\theta(\rho_0/\rho)^{2/3}}^{\theta_0} E(x,\rho_0)\frac{dx}{x^2}$$

$$+ \frac{\theta}{\theta_0} F(\theta_0,\rho_0). \tag{B.11}$$

If one follows the line $\rho = \rho_0(\theta/\theta_0)^{3/2}$ in the $\rho$–$\theta$ plane, the second term vanishes and (B.11) coincides with (3.19).

It is also worth remarking that all those formulae hold for the high-temperature phase as well if one puts $\bar{g} = g$.

### Appendix C. Perturbation expansions

In I, Eq. (5.2) was being solved with the help of expansion in powers of the Fourier transform of the potential $U_2(r)$. However, $U_2(r)$ itself is an unknown function determined by other equations of the hierarchy. In the present appendix we shall show that one can look for a solution of the equations of the hierarchy in terms of expansions in powers of the interaction potential $K(r)$, which permits us to answer some general questions. For convenience we introduce a dimensionless parameter $\lambda$, so that $K(r) = \lambda \tilde{K}(r)$, and shall speak of expansions in powers of $\lambda$.

We shall assume that the external potential $V^{(e)} = 0$. By making use of (I.C.1) we recast (2.9) in the form

$$\nabla_1 U_s(\mathbf{x}_s) = \nabla_1 \sum_{j=2}^{s} K\left(\left|\mathbf{r}_1 - \mathbf{r}_j\right|\right) + \frac{\int \rho_{s+1}(\mathbf{x}_{s+1}) \nabla_1 K\left(\left|\mathbf{r}_1 - \mathbf{r}_{s+1}\right|\right) d\mathbf{r}_{s+1}}{\dfrac{1}{N-s} \int \rho_{s+1}(\mathbf{x}_{s+1}) d\mathbf{r}_{s+1}}. \qquad (C.1)$$

Let us choose a value of $s$ and find $\rho_{s+1}$ upon setting $U_{s+1} = 0$. To this end we observe that then Eq. (2.21) gives $u_{s+1}(\mathbf{x}_{s+1}) = 1$ according to the limiting condition, while (2.8) yields

$$v_{s+1}(\mathbf{x}_{s+1}, \mathbf{m}_{s+1}, z) = \left(z - \frac{1}{2m} \sum_{j=1}^{s+1} \mathbf{p}_j^2\right)^{-1}. \qquad (C.2)$$

Now by (2.28) we compute $\rho_{s+1}(\mathbf{x}_{s+1})$ in this case, which allows us to calculate the term of $U_s$ proportional to $\lambda$ with the aid of (C.1). It is to be emphasized that this term will be known exactly.

Next, we rewrite (2.8) as

$$\frac{\hbar^2}{2m} \sum_{j=1}^{s} \nabla_j^2 v_s + \frac{i\hbar}{m} \sum_{j=1}^{s} \mathbf{p}_j \nabla_j v_s + \left(z - \frac{1}{2m} \sum_{j=1}^{s} \mathbf{p}_j^2\right) v_s = 1 + U_s(\mathbf{x}_s) v_s. \qquad (C.3)$$

Once the term of order $\lambda$ in $U_s$ is known, this equation permits us to find terms of orders $\lambda^0$ and $\lambda$ in $v_s$. In like fashion, upon reformulating (2.21) as

$$\frac{\hbar^2}{2m} \sum_{j=1}^{s} \nabla_j^2 u_s(\mathbf{x}_s) + \frac{i\hbar \mathbf{p}_0}{m} \sum_{j=1}^{s} \nabla_j u_s(\mathbf{x}_s) = U_s(\mathbf{x}_s) u_s(\mathbf{x}_s), \qquad (C.4)$$

we can compute terms of orders $\lambda^0$ and $\lambda$ in $u_s$ (the first of them is 1). Now we are able to calculate $\rho_s(\mathbf{x}_s)$, too, by (2.28) to the same order in $\lambda$.

We substitute the expression obtained for $\rho_s(\mathbf{x}_s)$ into (C.1) written for $U_{s-1}$ and calculate terms of orders $\lambda$ and $\lambda^2$ in $U_{s-1}$, which will give terms of orders $\lambda^0$, $\lambda$ and $\lambda^2$ in $v_{s-1}$, $u_{s-1}$ and $\rho_{s-1}$ by (C.3), (C.4) and (2.28). Proceeding in the same manner we shall finally arrive at $g(\mathbf{r}) = \rho_2(\mathbf{r})/\rho^2$ for which we shall know terms in $\lambda^0, \lambda, ..., \lambda^{s-1}$. As to thermodynamic quantities, we shall know terms in $\lambda^0, \lambda, ..., \lambda^s$ according to (3.8) and (3.16) because Eq. (3.17) too can be solved by expanding in powers of $\lambda$ analogously with Sec. 7 of I. It is important to stress that all those terms will be known exactly. Hence, by starting from Eq. (C.1) with an appropriate number $s$ we are able to calculate any term in the perturbation expansion of any quantity. It should be observed that the present procedure holds also for the equations of I because only Eq. (C.4) is absent there.

The method considered proves the existence of a solution to the hierarchy of the present paper as well as to the hierarchy of I. Of course, it is yet necessary to prove that the relevant series converge, which is not a simple matter. As is usual with perturbation expansions we suppose that they converge (perhaps, asymptotically) at least for not too large $\lambda$, i.e., in the case of a weak coupling.

The result obtained shows also that the hierarchy of equations can be solved without any approximation introduced usually, when one deals with a hierarchy, to cut off it. An analogous result exists for the classical BBGKY hierarchy as well [19]. In practice, however, one is obliged to introduce some approximation or other for the resulting series prove to be very complex and impractical to be handled.

The present method enables one also to demonstrate the uniqueness of the solution. The functions $\rho_s$ and $U_s$ are determined uniquely by Eqs. (2.28) and (C.1) since the constant appearing when integrating (C.1) is found from the condition that $U_s = 0$ at infinity (see I). So we need consider only (C.3) and (C.4). We observe first that the left side of (C.4) is a particular case of the left side of (C.3) if one puts $\mathbf{p}_i = \mathbf{p}_0$ for all $i$ and $z = s\, p_0^2/2m$. Therefore it is sufficient to discuss (C.3) alone. If the right side is given, Eq. (C.3) is a linear differential equation with constant coefficients. Therefore it should have a unique solution subject to the conditions formulated at the end of Sec. 2 of I.

The above argument shows that the solution of the hierarchy in terms of perturbation expansions is unique. From this it follows an important consequence. Recall that we presume the potential $K(\mathbf{r})$ to be spherically symmetric and imply homogeneous fluids. When we put $U_{s+1} = 0$, we obtain that $u_{s+1}(\mathbf{x}_{s+1}) = 1$ while $v_{s+1}$ is given by (C.2). These functions as well as functions calculated by (2.28), (C.1), (C.3) and (C.4) will describe an isotropic system, be it in the high-temperature or low-temperature state. Hence, the unique solution obtained by the procedure of the present appendix will always correspond to an isotropy in space. This being so, by (3.5) and (3.15) we shall definitively get $\mathbf{p}_0 = 0$, that is to say, we shall have a condensate phase without superfluidity. This last conclusion, however, extends only to week potentials $K(\mathbf{r})$ when one may expect the convergence of the relevant series. In case the potential is not week, the uniqueness may fail and a superfluid solution for which $\mathbf{p}_0 \neq 0$ may emerge.

## Appendix D. Derivation of Eq. (6.3)

Instead of (I.6.1), we assume the potential

$$K(r) = \begin{cases} K_m, & \text{if } r < a \\ 0, & \text{if } r > a. \end{cases} \qquad (D.1)$$

We consider spherically symmetric solutions. Upon putting $U_2(r) = K(r)$ in (2.24) and writing $u_2(r) = \chi(r)/r$ we get

$$\frac{d^2\chi}{dr^2} - \frac{m}{\hbar^2} K(r)\chi(r) = 0. \qquad (D.2)$$

As long as $u_2 \to 1$ as $r \to \infty$, this equation yields, for $r > a$,

$$u_2(r) = 1 - \frac{C_1}{r}. \qquad (D.3)$$

If $r < a$, Eq. (D.2) leads to the following expression for $u_2(r)$ finite at $r = 0$

$$u_2(r) = \frac{C_2}{r} \sinh br, \qquad b^2 = \frac{m}{\hbar^2} K_m. \qquad (D.4)$$

As usual the constants $C_1$ and $C_2$ are specified by the condition that $u_2(r)$ and its derivative shall be continuous at $r = a$. As a result, when $r \leq a$,

$$u_2(r) = \frac{\sinh br}{br \cosh ba}. \qquad (D.5)$$

We consider now an integral of the type

$$\int_0^\infty f(r) K(r) \frac{\partial g_c}{\partial r} dr = K_m \frac{\rho_c}{\rho} \int_0^a f(r) \frac{du_2^2}{dr} dr \qquad (D.6)$$

with an arbitrary function $f(r)$ assumed to be smooth. On the right, we have made use of (D.1) and replaced $g_c(r)$ by the first term of (5.1). From (D.5) it follows that, at $r \leq a$ except $r = a$, the function $u_2(r)$ is exponentially small when $b \to \infty$. For this reason, only the value of $f(r)$ at $r = a$ is of importance for the integral in (D.6). Upon putting $f(r) = f(a)$, the integral is easily evaluated and leads to (6.3) in the limit as $K_m \to \infty$.

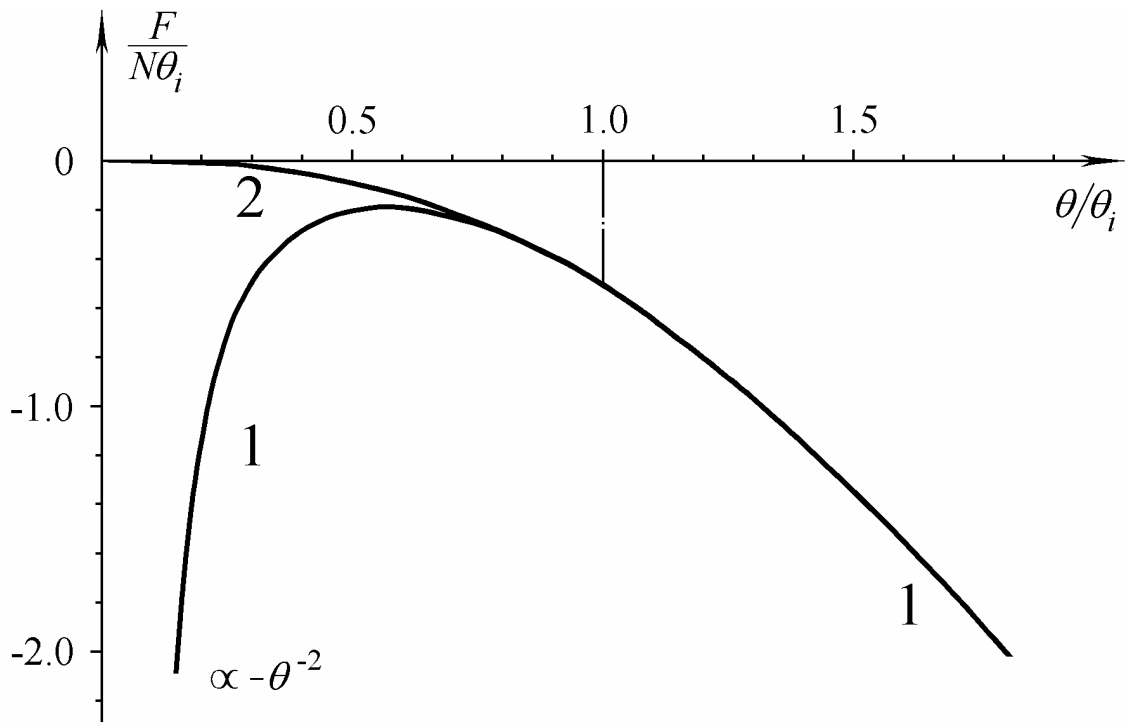

**Figure 1**. Dependence of the free energy $F$ of an ideal Bose gas on the temperature $\theta$ at a given $\rho$. The condensation temperature is $\theta_i = (\rho/\zeta_3\omega)^{2/3}$, the constants $\zeta_k$ being defined in I ($\zeta_3 = 2.315157$, $\zeta_5 = 1.783293$).

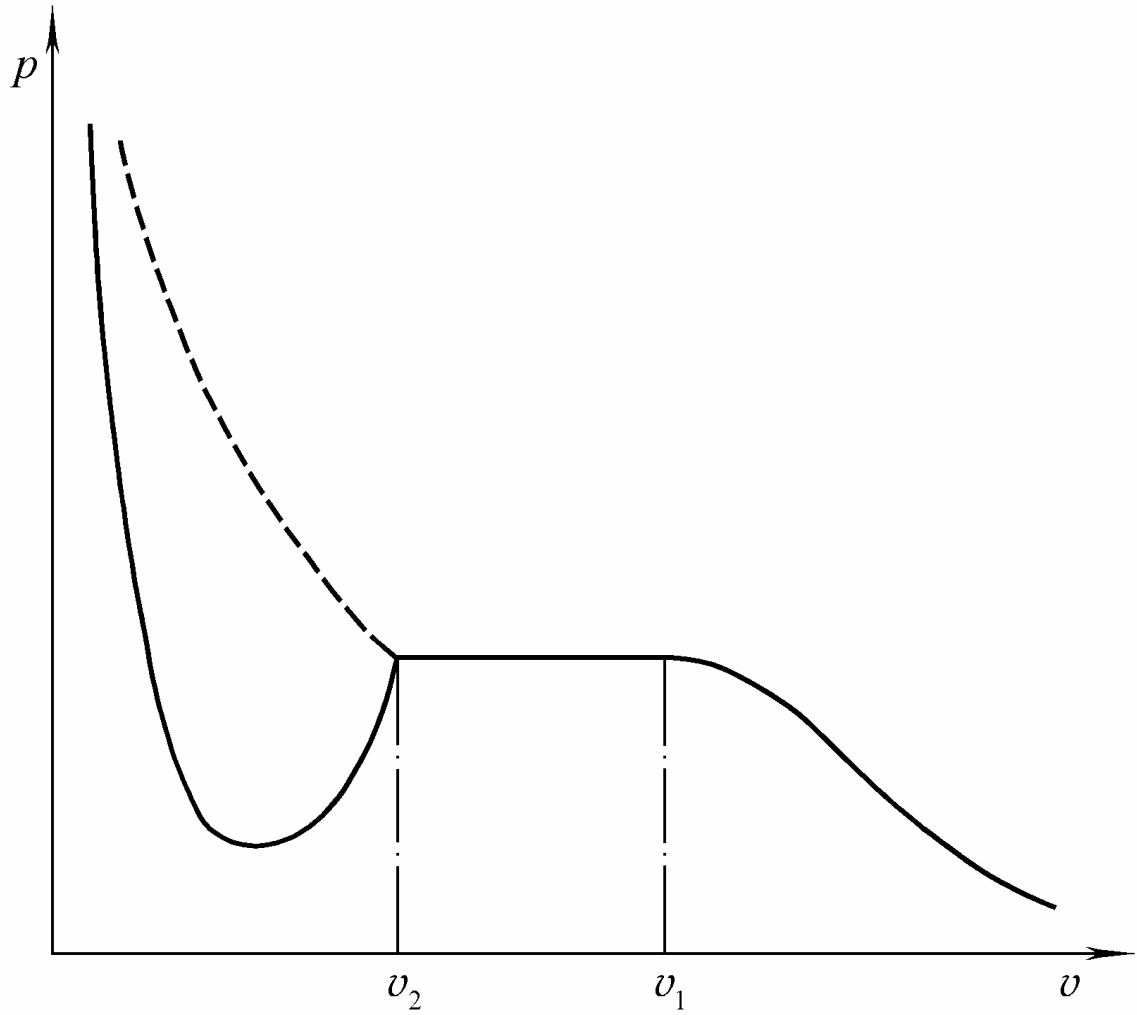

**Figure 2**. Isotherms of a hard-sphere Bose system under the neglect of triplet correlations. The specific volumes $v_1 = 1/\rho_1$ and $v_2 = 1/\rho_2$ are given by (6.16) and (6.25) respectively.